\documentstyle[onecolumn]{mn}
\input{psfig}
\input{epsf.tex}
\newcommand{\mincir}{\raise -2.truept\hbox{\rlap{\hbox{$\sim$}}\raise5.truept
\hbox{$<$}\ }}
\newcommand{\magcir}{\raise -2.truept\hbox{\rlap{\hbox{$\sim$}}\raise5.truept
\hbox{$>$}\ }}
\newcommand{\minmag}{\raise-2.truept\hbox{\rlap{\hbox{$<$}}\raise 6.truept\hbox
{$>$}\ }}
\newcommand{\be}{\begin{equation}}
\newcommand{\ee}{\end{equation}}
\newcommand{\ba}{\begin{eqnarray}}

\newcommand{\ea}{\end{eqnarray}}
\newcommand{\brr}{\begin{array}}
\newcommand{\err}{\end{array}}
\newcommand{\bc}{\begin{center}}
\newcommand{\ec}{\end{center}}
\newcommand{\br}{\mbox{\bf r}}

\newcommand{\bS}{\mbox{\bf S}}
\newcommand{\bq}{\mbox{\bf q}}
\newcommand{\bx}{\mbox{\bf x}}

\newcommand{\bk}{\mbox{\bf k}}
\newcommand{\rmb}{\mbox{\rm  b}}

\newcommand{\om}{\omega}
\newcommand{\etal}{{et al.}~}

\newcommand{\p}{\partial}
\newcommand{\f}{\frac}

\newcommand{\Om}{\Omega}
\newcommand{\w}{\omega}
\newcommand{\de}{\delta}
\newcommand{\ded}{\delta_{_D}}
\newcommand{\eps}{\epsilon}
\newcommand{\s}{\sigma}
\newcommand{\al}{\alpha}
\newcommand{\lam}{\lambda}
\newcommand{\fde}{\tilde{\delta}}

\newcommand{\fW}{\widetilde{W}}
\newcommand{\bfx}{{\bf x}}

\newcommand{\bfv}{{\bf v}}

\newcommand{\bfq}{{\bf q}}

\newcommand{\bfu}{{\bf u}}

\newcommand{\calN}{{\cal N}}
\newcommand{\calH}{{\cal H}}

\newcommand{\lan}{\langle}
\newcommand{\ran}{\rangle}
\newcommand{\epsm}{\epsilon_{_{M}}}
\newcommand{\epsmo}{\epsilon_{_{M_0}}}
\newcommand{\epsmu}{\epsilon_{_{M_1}}}
\newcommand{\epsmt}{\epsilon_{_{M_2}}}

\newcommand{\sigm}{\sigma_{_{\!M}}}
\newcommand{\sigmu}{\sigma_{_{\!M_1}}}
\newcommand{\sigmt}{\sigma_{_{\!M_2}}}

\newcommand{\0}{\circ}
\newcommand{\nps}{n_{_{\!{P\!S}}}}

\title[Halo bias field]
{The bias field of dark matter haloes} 
\author[P. Catelan, F. Lucchin, S. Matarrese and C. Porciani]
{ {\Large Paolo Catelan$^{1,2}$, Francesco Lucchin$^{3}$, 
Sabino Matarrese$^{4}$
and Cristiano Porciani$^{5}$} \\ 
$^1$ Theoretical Astrophysics Center, Juliane Maries Vej 30, DK-2100 
Copenhagen $\O$, Denmark\\
$^2$ Department of Physics, Astrophysics, Nuclear Physics Laboratory,
Keble Road OX1 3RH, Oxford, UK \\
$^3$ Dipartimento di Astronomia, Universit\`{a} di Padova, vicolo
dell'Osservatorio 5, I--35122 Padova, Italy\\
$^4$ Dipartimento di Fisica {\em Galileo Galilei}, Universit\`{a} di
Padova, via Marzolo 8, I--35131 Padova, Italy\\
$^5$ SISSA, Scuola Internazionale di Studi Superiori Avanzati,
via Beirut 2-4, I--34014 Trieste, Italy}

\begin{document}

\maketitle

\begin{abstract}
This paper presents a stochastic approach to the clustering evolution
of dark matter haloes in the Universe. Haloes, identified by a
Press-Schechter-type algorithm in Lagrangian space, are described in
terms of `counting fields', acting as non-linear operators on the
underlying Gaussian density fluctuations.  By ensemble averaging these
counting fields, the standard Press-Schechter mass function as well as
analytic expressions for the halo correlation function and
corresponding bias factors of linear theory are obtained, extending the
recent results by Mo and White.  The non-linear evolution of our halo
population is then followed by solving the continuity equation, under
the sole hypothesis that haloes move by the action of gravity. This
leads to an exact and general formula for the {\em bias field~} of dark
matter haloes, defined as the local ratio between their number density
contrast and the mass density fluctuation.  Besides being a function of
position and `observation' redshift, this random field depends upon the
mass and formation epoch of the objects and is both non-linear and
non-local. The latter features are expected to leave a detectable
imprint on the spatial clustering of galaxies, as described, for
instance, by statistics like the bispectrum and the skewness. Our
algorithm may have several interesting applications, among which the
possibility of generating mock halo catalogues from low-resolution
N-body simulations.

\end{abstract}


\section{Introduction}

The theory proposed by Press and Schechter (1974, hereafter PS) to
obtain the relative abundance of matter condensations in the Universe
has strongly influenced all later studies on the statistical properties
of dark matter haloes and led to a large variety of extensions,
improvements and applications. Actually, already in the sixties,
Doroshkevich (1967) had derived the mass distribution function for
`newly generated cosmic objects', completely analogous to the PS one;
he had also clearly pointed out the existence of what has been later
referred to as {\em cloud-in-cloud problem} (e.g. Bardeen \etal
1986). The `Press-Schechter model', which is based on the gravitational
instability hypothesis, is now considered as one of the cornerstones of
the hierarchical scenario for structure formation in the Universe. It
shows, in fact, how gravitational instability makes more and more
massive condensations grow by the aggregation of smaller units, only
provided the initial density fluctuation field contains enough power on
small scales. The main drawback of the original PS model is indeed the
cloud-in-cloud problem, i.e. the fact that their procedure selects
bound objects of given mass that can have been already included in
larger mass condensations of the same catalog.  The problem was later
solved by several authors (Peacock \& Heavens 1990; Bond \etal 1991;
Cole 1991) according to the so-called `excursion set' approach, by
calculating the distribution of first-passage `times' across the
collapse threshold for suitably defined random walks.  Lacey and Cole
(1993, 1994) implemented these ideas to study the merger rates of
virialized haloes in hierarchical models of structure formation.

An important aspect of the PS model is that, being entirely based on
linear theory, suitably extrapolated to the collapse time of spherical
perturbations, it is, by definition, local in Lagrangian space. While
this Lagrangian aspect of the theory does not have immediate
implications for the study of the mass function of dark matter haloes,
it is, instead, of crucial importance for their spatial clustering
properties. This point was recognized by Cole and Kaiser (1989) and,
more recently, by Mo and White (1996, hereafter MW), who proposed a
bias model for halo clustering in Eulerian space, by a suitable
extension of the original PS algorithm for the mass function. With
their formalism MW studied the clustering of dark matter haloes with
different formation epochs (see also Mo, Jing \& White 1996). The
comparison of their theoretical predictions with the spatial
distribution of haloes obtained by a {\em friends-of-friends} group
finder and by a spherical overdensity criterion in numerical
simulations proved extremely successful.

These very facts imply that there exists a local version of the PS
algorithm providing a {\em mapping~} between points of Lagrangian space
and the haloes in embryo which will come into existence at the various
epochs.  For a given realization of the initial density field, the PS
mapping is such that, at a fixed redshift $z$, each Lagrangian point
$\bq$ can be assigned to a matter clump of some mass $M$, identified by
a suitable Lagrangian filter, which is collapsing at the epoch $z_f=z$.
One can therefore exploit the existence of this mapping to assign a
stochastic halo process, our {\em halo counting field~} below, to each
point $\bq$. This will be the starting point of our analysis.

What the PS ansatz cannot account for is the fact that the fluid
elements are moved apart by gravity, so that the halo which the PS
mapping assigns to the fluid patch with Lagrangian coordinate $\bq$ is
not going to collapse in the same position, i.e. at $\bx=\bq$, but,
rather in the Eulerian point $\bx(\bq,z)=\bq + \bS(\bq,z)$, with
$\bS(\bq,z)$ the displacement vector, corresponding to the Lagrangian
one at the epoch $z=z_f(\bq,M)$ of its collapse.  This fact, while not
affecting in any way the PS result for the mean mass function, as the
average halo abundance cannot change by scrambling the objects,
sensibly modifies their spatial clustering properties. Modeling the
latter effect is one of the main purposes of the present work.  In
their derivation of the Eulerian halo bias MW took into some account
this problem by allowing for the local compression, or expansion, of
the volumes where the haloes are located, an effect which is of crucial
importance for the derivation of the correct halo density contrast.
Their derivation, however, is formally flawed by the fact that they
only deal with mean halo number densities, so that they are forced to
define the bias in terms of them. For reasons to be shown below,
however, this heuristic treatment can be put on sounder statistical
grounds, by applying a suitable coarse-graining procedure.

Of course, the PS model has its own limitations. The comparison of its
predictions for the mass function with the outputs of N-body
simulations (e.g.  Efstathiou \etal 1988; Gelb \& Bertschinger 1994;
Lacey \& Cole 1994), while surprisingly successful in its general
trends, given the simplicity of the assumptions, showed a number of
problems. Gelb and Bertschinger (1994), for instance, found that the
simulated haloes are generally less massive than predicted, the reason
being that merging does not erase substructure in large haloes as fast
as required by the PS recipe.

There have been many attempts to improve the original PS model. If
cosmic structures preferentially formed at the peaks of the initial
density fluctuation field this would affect their mean mass function
(Bardeen \etal 1986; Bond 1988; Colafrancesco, Lucchin \& Matarrese
1989; Peacock \& Heavens 1990; Manrique \& Salvador--Sol\'e 1995,
1996).  Bond and Myers (1996) developed a {\em peak-patch} picture of
cosmic structure formation, according to which virialized objects are
identified with suitable peaks of the Lagrangian density field. The
peak-patch collapse dynamics is then followed in terms of the
homogeneous ellipsoid model, which allows for the influence of the
external tidal field, while the Zel'dovich approximation (Zel'dovich
1970) is used for the external peak-patch dynamics.  The effects of
non-spherical collapse on the shape of the mass distribution were
studied by Monaco (1995).  Lee and Shandarin (1997) analytically
derived the mass function of gravitationally bound objects in the frame
of the Zel'dovich approximation.

We prefer here to follow the simple lines of the PS theory to set up
the `initial conditions' for our stochastic approach to the evolution
of halo clustering. Nevertheless, one should keep in mind that our
approach is flexible enough to accept many levels of improvement in the
treatment of the Lagrangian initial conditions.

A relevant part of the following analysis will be devoted to the study
of the evolution of halo clustering away from the linear regime. It
turns out that the problem can be solved exactly in terms of the
evolved mass density.  An important result of this analysis is that the
general Eulerian bias factor, defined as the local ratio between the
halo density contrast and the mass fluctuation field, is both
non-linear and non-local. The latter property follows directly from our
selection criterion of candidate haloes out of the linear density
field.

Our algorithm can also be seen as a specific example of a bias model
which is local in Lagrangian space. This is expected to have relevant
consequences on galaxy clustering.  Because of this local Lagrangian
character, our model strongly differs from the local Eulerian bias
prescription applied by Fry and Gazta\~naga (1993) to the analysis of
the hierarchical correlation functions.  A simple test of our theory
can be obtained by analyzing the behaviour of the bispectrum (or the
skewness), whose shape (scale) dependence will be shown to be directly
sensitive to the assumption of local bias in Lagrangian vs.  Eulerian
space.

Our results for the evolved halo distribution generally allow to study
their statistical properties at the required level of non-linearity,
and could be further implemented to generate mock halo catalogues
starting from low-resolution numerical simulations of the
dissipationless matter component.  These results have important
implications for the study of the redshift evolution of galaxy
clustering, a problem made of compelling relevance by the growing body
of observational data at high-redshift which are being produced by the
new generation of large telescopes.  A general study of this problem
has been recently performed by Peacock (1997) and Matarrese \etal
(1997); the latter pointed out that knowledge of the evolution of the
effective bias for the various classes of objects is a key ingredient
in the comparison of theoretical scenarios of structure formation with
observational data on clustering at high redshift. Kauffmann, Nusser
and Steinmetz (1997) used both semi-analytical methods and N-body
techniques to study the physical origin of bias in galaxies of
different luminosity and morphology.

The plan of the paper is as follows. In Section 2 we define our halo
counting field, within the linear approximation, both in the Lagrangian
and Eulerian context. The non-linear evolution of the halo clustering
is analyzed in Section 3, where we also compute the bispectrum and
skewness of the evolved halo distribution. Section 4 contains a general
discussion of our results and some conclusions.

\section{Stochastic approach to halo counting and clustering}

\subsection{Basic tools and notation}

Let us assume that the mass density contrast $\eps(\bq)$, linearly
extrapolated to the present time, is a statistically homogeneous and
isotropic Gaussian random field completely determined by its
power-spectrum $P(k)$. Here $\bq$ represents a comoving Lagrangian
coordinate. A smoothed version of the field $\eps(\bq)$ is obtained by
convolving it with a rotationally invariant filter $W_R(q)$, containing
a resolution scale $R$, with associated mass $M\sim \rho_b\,R^3$,
$\rho_b$ being the background mean density at $z=0$,
\be
\eps_{_R}(\bq)=\int 
d\bq'\, W_R(|\bq-\bq'|)\,\eps(\bq')\equiv \epsm\!(\bq)\;.
\label{}\ee
The smoothed field is also Gaussian with one--point distribution
function
$G_{\sigm}\!(\epsm)=(2\pi\,\sigm^2)^{-1/2}\,\exp(-\epsm^2/2\sigm^2)$,
where $\sigm^2$ denotes the variance of $\epsm$,
$\sigm^2\equiv\lan\epsm^2\ran=(2\pi^2)^{-1}\int_0^\infty dk\,k^2\,P(k)
\widetilde W(kR)^2\,.$ The symbol $\fW(kR)$ indicates the Fourier
transform of the filter function.  In the following, we will often be
concerned with the joint distribution of the fields $\epsmu\!(\bq)$ and
$\epsmt\!(\bq)$.  The two--point correlation function of the linear
density contrast smoothed on the scale $R_1$ and $R_2$ is
\be 
\xi_{12}(q)= \lan
\,\epsmu\!(\bq_1) \, \epsmt\!(\bq_2) \, \ran = \f{1}{2
\pi^2}\int_0^\infty dk\,k^2\, P(k) 
\,\fW(kR_1) \,\fW(kR_2)
\,j_\0(kq)\;, 
\label{eq:corr}\ee 
where $q=|\bq_1-\bq_2|$ and $j_\0(x)$ is the spherical Bessel function
of order zero.  We term $\s_{12}^2$ the value assumed by $\xi_{12}$ in
the limit $q \to 0$.

The properties of the filtered quantities clearly depend upon the
choice of the window function. For instance, the relation between the
mass enclosed by a top-hat filter $W_R(q)=3\,\Theta(R-q)/4\pi R^3$
(where $\Theta(x)$ is the Heaviside step function) is the standard
$M(R)=4\pi\rho_b R^3/3$. Instead, for a Gaussian window, $W_R(q)=(2\pi
R^2)^{-3/2}\exp(-q^2/2R^2)$, the enclosed mass is
$M(R)=(2\pi)^{3/2}\rho_b R^3$. These two masses coincide for $R_G =
0.64\,R_{TH}$ (Bardeen \etal 1986).

In the literature, the sharp top-hat filtering has been alternatively
adopted in Fourier space, $\fW_R(k)= \Theta (k_R-k)$, where $k_R=1/R$
and $k=|\bk|$.  The most remarkable property of this filter is that
each decrease of the smoothing radius adds up a new set of uncorrelated
modes (Bardeen \etal 1986; Bond \etal 1991; Lacey \& Cole 1993). This
also implies that, for example, the correlation function in
eq.(\ref{eq:corr}) simplifies to $\xi_{12}= \xi_{11}$, whenever
$k_{R_1} < k_{R_2}$; consequently, $\s_{12}=\s_{11}\equiv \s_1$. In
practice, the information is always erased below the largest of the two
smoothing lengths.  This property will be particularly useful in the
next sections. For this `sharp k-space' filter, the main difficulty is
how to associate a mass $M(R)$ to the cutoff wavenumber $k_R$. Lacey and
Cole (1993) give the expression $M(R)=6\pi^2\rho_b k_R^{-3}$, which
coincides with the mass within a top-hat filter if one takes
$k_R=2.42/R_{TH}$.

In the next section we introduce the halo counting random fields that
allow a fully stochastic description of the biased haloes
distribution.  To illustrate how our formalism works, we first show how
to derive the PS mass function by performing a simple averaging of our
stochastic counts.

\subsection{Lagrangian mass function: Press-Schechter theory}

Press and Schechter proposed a simple model to compute the comoving
number density of collapsed haloes directly from the statistical
properties of the linear density field, assumed to be Gaussian.
According to the PS theory, a patch of fluid is part of a collapsed
region of scale larger than $M(R)$ if the value of the smoothed linear
density contrast on that scale exceeds a suitable threshold $t_f$.  The
idea is to use a global threshold in order to mimic non-linear
dynamical effects ending up with halo collapse and virialization.  An
exact value for $t_f$ can be obtained by describing the evolution of
the density perturbations according to the spherical top-hat model. In
this case, a fluctuation of amplitude $\eps$ will collapse at the
redshift $z_f$ such that $\eps(\bq) = t_f\equiv\de_c/D(z_f)$, where
$D(z)$ denotes the linear growth factor of density perturbations
normalized as $D(0)=1$. In the Einstein-de Sitter universe and during
the matter dominated era the critical value $\de_c$ does not depend on
any cosmological parameter and is given by $\de_c=3(12 \pi)^{2/3}/20
\simeq 1.686$, while, for general non-flat geometries, its value shows
a weak dependence on the density parameter $\Omega$, the cosmological
constant $\Lambda$ and the Hubble constant $H$ (e.g. Lacey \& Cole
1993), thus on redshift. In a flat universe with vanishing cosmological
constant $D(z)=(1+z)^{-1}$; explicit expressions for the linear growth
factor are given in Appendix A for general Friedmann models.

A local version of the PS approach can be built up in terms of
stochastic counting operators acting on the underlying Gaussian density
field, as follows. The number of haloes per unit mass, contained in the
unit comoving volume centered in $\bq$, identified by the
collapse-threshold $t_f(z_f)$, is described as a density field of a
point process by
\be
\calN_h^L(\bq|M,t_f)=-2 \,\f{\rho_b}{M} \, \f{\p}{\p M}
\Theta\big[\epsm\!(\bq)-t_f\big]\;.
\label{eq:defNHL}\ee
Note that the quantity $\calN_h^L(\bq|M,t_f)$ is non-zero only when the
filtered density contrast in $\bq$ upcrosses, or downcrosses, the
threshold $t_f$, by varying the smoothing length $R$ (or the
corresponding mass $M$). The factor of 2, appearing in the expression
of $\calN_h^L(\bq|M,t_f)$, is needed in order to obtain the right
normalization of the mass function, in which case it has been shown to
be intimately related to the solution of the cloud-in-cloud problem
(Peacock \& Heavens 1990; Bond \etal 1991; Cole 1991), at least for
sharp k-space filtering.  At this level, our description should be
thought of as a sort of differential version of Kaiser's bias model
(Kaiser 1984), that defines a population of objects with the right
average halo abundance and their related clustering properties, rather
than a detailed modeling of how structures form from the primordial
density field.  In a forthcoming paper, however, we will show that the
present approach is fully consistent with a rigorous treatment of the
cloud-in-cloud problem (Porciani \etal 1997). In that approach halo
correlations will be obtained from pairs of first-upcrossing `times'
for spatially correlated random walks above the collapse threshold
$t_f$.

It can be seen from equation (\ref{eq:defNHL}) that a population of
haloes is uniquely specified by the two parameters $M$ and $t_f$. In
the standard PS formulation $t_f$ is interpreted as a sort of time
variable, related to the formation redshift $z_f$, which decreases with
real time, as every halo continuously accretes matter. In this sense
one can say that, for a continuous density field with infinite mass
resolution, each halo disappears as soon as it forms to originate
another halo of larger mass.

Alternatively, instead of considering $t_f$ as a time variable, one can
use it simply as a label attached to each halo. The haloes so labelled
can be thought as keeping their identity during the subsequent
evolution at any observation redshift $z$. This is not in contrast with
the fact that in the real Universe dark matter haloes undergo merging
at some finite rate (e.g. Lacey \& Cole, 1993, 1994). Within such a
picture, in fact, the physical processes of accretion and merging
reduce to the trivial statement that haloes identified by a given
threshold are necessarily included in catalogues of lower threshold, so
that, in the limit of infinite mass resolution, only haloes with
$z_f=z$ would actually survive. Nevertheless, keeping $z_f$ distinct
from $z$ may have several advantages, among which the possibility of
allowing for a more realistic description of galaxy and cluster
formation inside haloes, for both the evolution of the luminosity
function (Cavaliere, Colafrancesco \& Menci 1993; Manrique \&
Salvador--Sol\'e 1996) and of the galaxy bias (e.g. MW; Kauffmann \etal
1997).  Let us stress, however, that we are not addressing here the
issue of galaxy or cluster merging: our method is completely general in
this respect and allows to span all possible models, from the
instantaneous merging hypothesis ($z_f=z$) to the case of no merging at
all ($z_f$ fixed for changing $z\leq z_f$).

In what follows, therefore, we will assume that we can deal with the
halo population specified by the formation threshold $t_f$ at any
redshift $z$. Only in this sense we will say that we `ignore' the
effects of merging in our description: merging can be exactly recovered
at any step, and with any assumed mass resolution, as the relation
between $z_f$ and $z$.  To implement this idea it is enough to scale
appropriately the argument of the Heaviside function in
eq.(\ref{eq:defNHL}), which can be recast in the form
\be
\calN_h^L(\bq,z|M, z_f)=
-2 \,\f{\rho_b}{M} \, \f{\p}{\p M}
\Theta\big[\epsm\!(\bq, z)-\de_f(z, z_f)\big]
 \;,
\label{eq:defNHL1}\ee
where $\epsm(\bq,z)\equiv D(z)\eps(\bq)$ and $\de_f(z, z_f)\equiv
\de_cD(z)/D(z_f)$.  It can be easily seen that the ensemble average of
the counting field $\calN_h^L(\bq,z|M,z_f)$ corresponds to the PS mass
function
\be
\lan\calN_h^L(\bq,z|M,z_f)\ran\,dM=\nps(z|M,z_f)\,dM\;,
\ee
where
\be
\nps(z|M,z_f)\,dM \equiv
\f{1}{\sqrt{2\pi}}\,\f{\rho_b}{M}\,
\f{\de_f(z, z_f)}{\sigm^3\!(z)}\,
\exp\left[ -\f{\de^2_f(z, z_f)}{2\sigm^2\!(z)}
\right]\,
\left|
\f{d\sigm^2\!(z)}{dM}
\right|\,dM \;.
\ee
Note that we emphasized the $z$-dependence of the comoving mass
function, though it is straightforward to verify that the value of
$\nps(z|M,z_f)$ does not change with $z$. In fact, since we are
ignoring the effects of merging, once a class of haloes has been
identified, their mean comoving density keeps constant in time. Thus,
as far as the mass function is concerned, the introduction of the
observation redshift $z$ is somewhat more formal than physical.
However, this distinction will be far more significant in the next
sections, where, in order to compute the halo-to-mass bias factor, we
will relate the Lagrangian distribution of a population of haloes
selected at $z_f$ to the mass density fluctuation field linearly
extrapolated to the redshift $z$. Models of galaxy formation which
assume that galaxies form at a given redshift $z_f$ with some initial
bias factor and that their subsequent motion is purely caused by
gravity (e. g. Dekel 1986; Dekel \& Rees 1987; Nusser \& Davis 1994;
Fry 1996) can be easily accommodated into this scheme.

To conclude this section, let us consider the integral stochastic process 
\be
\int_M^\infty dM' M'\calN_h^L(\bq,z|M',z_f)
= 2 \,\rho_b\, \Theta\big[\epsm\!(\bq,z)-\de_f(z,z_f)\big] \;,
\label{eq:kaiser}
\ee
representing the fraction of mass, in the unit Lagrangian comoving
volume centered in $\bq$, which at redshift $z_f$ has formed haloes
more massive than $M$. This coincides with the original Kaiser bias
model (Kaiser 1984) up to the multiplicative factor $2\rho_b$, which is
irrelevant for calculating correlation functions.

\subsection{Conditional Lagrangian mass function} 

The PS theory reviewed in the previous section describes the overall
distribution of halo masses in a homogeneous universe of mean density
$\rho_b$. However, of cosmological interest is also, for instance, the
estimate of the halo distribution within rich or poor environments
(which can be related to the galaxy number enhancement per unit mass in
rich clusters or in voids), thus justifying the investigation of the
distribution of halo masses conditioned to lie within a larger
uncollapsed container of given density. The conditional mass function
has been studied by several authors (e.g. Bond \etal 1991; Bower 1991;
Lacey \& Cole 1993).

We extend here the approach introduced in the previous section in order
to derive the conditional mass function. Specifically, we calculate the
comoving mass function, in the mass range $M$, $M+dM$, for objects
contained in a large region of dimension $R_\0$, corresponding to a
mass $M_\0$, with local density contrast $\eps_\0\equiv\epsmo$.  We
will require $\eps_\0\ll\delta_f$ and $R_\0\gg R$, to ensure that the
container is not collapsed yet by the epoch $z_f$ and that it encloses
a non-negligible population of objects.

In order to mimic these environmental effects, we modify the halo
counting field according to
\be 
\calN_h^L(\bq, z|M,z_f|M_\0, \eps_\0)=
-\f{2}{N_\0} \,\f{\rho_b}{M} \, 
\f{\p}{\p M}
\Theta\big[\epsm\!(\bq, z)-\de_f(z, z_f)\big] \,
\ded\!\big[\epsmo\!(\bq, z)-\eps_\0\big]\;, 
\label{eq:condfield}\ee 
where $\ded(\bq)$ denotes the Dirac delta function and $N_\0 \equiv
\lan\ded[\epsmo\!(\bq, z)-\eps_\0]\ran$ is the normalization
constant. Here the scalar $\eps_\0$ indicates the value of the random
field $\epsmo(\bq, z)$. Taking the ensemble average (and using the
cross-variance $\s_{ij}$ for a sharp k-space filter) one eventually
obtains
\be \lan\,\calN_h^L(\bq, z|M,z_f|M_\0, \eps_\0)\,\ran\,dM =
\nps(z|M,z_f|M_\0,\eps_\0)\,dM \;, 
\label{eq:9}\ee 
where 
\be
\nps(z|M,z_f|M_\0,\eps_\0)\,dM = \f{1}{\sqrt{2\pi}}\, \f{\rho_b}{M}
\f{\de_f(z,z_f)-\eps_\0 }{[\sigm^2\!(z)-\s^2_\0(z)]^{3/2}}\, \exp \left\{
-\f{[\de_f(z,z_f)-\eps_\0]^2}{2\,[\sigm^2\!(z)-\s^2_\0(z)]} \right\}\, \left|
\f{d\sigm^2\!(z)}{dM} \right|\,dM \;.  
\label{eq:conditps}\ee 
This straightforward calculation shows how to obtain results already
known in the literature by simply starting from the random field in
eq.(\ref{eq:condfield}): averaging that halo counting field leads to
the expected conditional mass function. But not only: unlike previous
treatments, once the halo counting field has been consistently defined,
other statistics, like the two-point halo correlation function, can be
calculated. We will carry out this program in the next section.

\subsection{Lagrangian clustering: halo-to-mass bias from correlations}

In this section we will compute the halo-halo correlation function
which coincides with the correlation function of our random counting
field. Specifically, we will calculate the Lagrangian halo correlation
function from the Lagrangian counting field $\calN_h^L(\bq,z|M,
z_f)$. By definition, the correlation function of this stochastic
process is given by
\be \xi_{hh}^L(q, z|M_1,z_1; M_2,z_2)=
\f{\lan\,\calN_h^L[\bq_1,z|M_1,\de_f(z, z_1)]
\;\calN_h^L[\bq_2,z|M_2,\de_f(z,z_2)] \, \ran} {\lan \,
\calN_h^L[\bq_1,z|M_1,\de_f(z,z_1)] \, \ran \, \lan\,
 \calN_h^L[\bq_2,z|M_2,\de_f(z,z_2)]\, \ran}-1\;,
\label{eq:corrhalo}\ee
where $q=|\bq_1-\bq_2|$.  Performing the ensemble average over the
Gaussian fields $\epsmu(\bq)$ and $\epsmt(\bq)$, we obtain
\be 
\lan\,\calN_h^L[\bq_1,z|M_1,\de_f(z,z_1)] \,
 \calN_h^L[\bq_2,z|M_2,\de_f(z,z_2)] \, \ran \, =\, 
\f{4\rho_b^2}{M_1 M_2}\, \f{\p}{\p{M_1}}\,\f{\p}{\p{M_2}}\,
\int_{\de_f(z, z_1)}^\infty \int_{\de_f(z, z_2)}^\infty d\alpha_1\, d\alpha_2
\,G_2(\alpha_1,\alpha_2) \;,
\label{eq:kash}
\ee 
where $G_2(\alpha_1,\alpha_2)$ denotes the bivariate Gaussian
distribution
\be 
G_2(\alpha_1,\alpha_2)=\left[ 2 \pi \s_1\s_2 \sqrt{
1-\om^2 }\right]^{-1} \, 
\exp{\left[ -\Big(\f{\alpha_1^2}{\s_1^2}+
\f{\alpha_2^2}{\s_2^2}-2\,\om\f{\alpha_1}{\s_1} \f{\alpha_2}{\s_2}
\Big)\Big/\! 2\big( 1-\om^2 \big)\right]}\;, 
 \ee 
with normalized correlation $\om(q)=\xi_{12}(q)/\sigmu\sigmt$ and
$\s_i\equiv D(z)\s_{_{\!M_i}}$.

The full exact expression for the halo-halo correlation function can be
obtained after an incredibly long algebraic computation. We report here
only the final expression.  Defining $\de_{fi}\equiv\de_f(z, z_i)$, we
have 
\ba 
1+\xi_{hh}^L(q,z|M_1,z_1;M_2,z_2) \, 
&=&\,\f{1}{\sqrt{1-\w^2}}
\left\{ \f{d\s_1}{dM_1}\,\f{d\s_2}{dM_2}+\f{\s_2^2}{\de_{f2}(1-\om^2)}
\,\left( \f{\de_{f1}}{\s_1}-\om \f{\de_{f2}}{\s_2} \right)
\,\f{d\s_1}{d M_1} \,\f{\p\om}{\p M_2} \, \right.  
\nonumber \\ 
&+&\,\f{\s_1^2}{\de_{f1}\,(1-\om^2)} \,
\left(\f{\de_{f2}}{\s_2}-\om\f{\de_{f1}}{\s_1}\right) \,\f{\p\om}{\p
M_1} \,\f{d \s_2}{d M_2} \, + \,\f{\s_1^2 \s_2^2}{\de_{f1} \de_{f2}} \,
\f{\p^2\om}{\p M_1\p M_2} 
\nonumber \\ 
&+&\,\f{\s_1^2\s_2^2}{\de_{f1}\de_{f2}(1-\om^2)^2}\, \left[\w(1-\w^2)+
(1+\om^2)\,\f{\de_{f1}}{\s_1}\,\f{\de_{f2}}{\s_2} \,-\,
\om\,\Big(\f{\de_{f1}^2}{\s_1^2}+\f{\de_{f2}^2}{\s_2^2}\Big)\right]\,
\nonumber \\ 
&\times& \,\f{\p\om}{\p {M_1}}\,\f{\p \om}{\p {M_2}} \bigg\}
\, \exp {\Bigg[ -\displaystyle{ \f{ \om^2 \Big( \displaystyle{
\f{\de_{f1}^2}{\s_1^2} } + \displaystyle{ \f{\de_{f2}^2}{\s_2^2}} \Big)
- \displaystyle{ 2\,\om\,\f{\de_{f1}}{\s_1} \f{\de_{f2}}{\s_2}}} {2
\left( 1- \om^2 \right)}} \Bigg]\, \left(\f{d \s_1}{d M_1} \f{d\s_2}{d
M_2}\right)^{-1} }\;.
\label{eq:k1}
\ea 
This expression can be easily shown to be independent of the
observation redshift $z$. A remark is now appropriate. Our formalism
describes the halo distribution as a discrete point process. However,
actual haloes are extended in size. This is why, as also seen in
numerical simulations, for separation smaller than the typical
Lagrangian radius of the halo, the correlation function abruptly
reaches the value $-1$: a sort of `exclusion principle' for extended
haloes.  Thus, we expect that the correlation function in
eq.(\ref{eq:k1}) can be a reliable description of halo clustering only
for $q ~\magcir {\rm Max} (R_1,R_2)$. Another point concerns the use of
finite mass resolution as in $N$-body simulations. The proper
analytical correlation to compare with in that case is the integral of
$\xi_{hh}^L\,\nps(M_1)\,\nps(M_2)$ over the specified mass interval,
appropriately normalized.

Since the action of the window functions on the correlations is
negligible for lags $q$ much larger than the smoothing lengths, $q \gg
R_1$ and $q \gg R_2$, for the normalized correlation we obtain
$\om(q)\simeq \xi_{m}(q)/\sigmu \sigmt$ [where $\xi_m(q)$ is the linear
mass autocorrelation function] and, eventually, for the halo
correlation
\be
\xi_{hh}^L(q, z| M_1, z_1; M_2, z_2) \, 
=  b_1^L(z|M_1,z_1)\,b_1^L(z|M_2,z_2)\,\xi_m(q,z) +
\f{1}{2}\,b_2^L(z|M_1,z_1)\,b_2^L(z|M_2,z_2)\,\xi^2_m(q,z) + \cdots\;. 
\label{eq:error}
\ee
Explicitly, the first two bias parameters read 
\be
b_1^L(z|M,z_f)=\f{\de_f(z, z_f)}{\sigm^2\!(z)}-\f{1}{\de_f(z, z_f)}=
\f{D(z_f)}{D(z)}\left[\f{\de_c}{D(z_f)^2\sigm^2}-\f{1}{\de_c}\right]\;,
\label{eq:b1}\ee 
\be 
b_2^L(z|M,z_f)=\f{1}{\sigm^2\!(z)} \left[ \f{\de^2_f(z, z_f)}{\sigm^2\!(z)}-3
\right] = \f{1}{D(z)^2\sigm^2} \left[ \f{\de_c^2}{D(z_f)^2\sigm^2}-3 \right]
\;.  
\label{eq:b2}\ee
These expressions for the bias factors generalize, in a sense, those
concerning the clustering properties of dark matter haloes in
Lagrangian space obtained by MW and Mo \etal (1996), with the relevant
difference that we have obtained the bias factor from the behaviour of
the halo two-point correlation function.  Also relevant is the fact
that, unlike MW, we obtained our Lagrangian correlation function
without introducing any background scale $R_\0$, which allows to extend
its validity down to spatial separation comparable to the halo size
$R\ll R_\0$. A calculation of the leading behaviour of the correlation
deriving from equations (\ref{eq:corrhalo}) and (\ref{eq:kash}) has
been already attempted by Kashlinsky (1987) who, however, missed the
contributions originated by the differentiation of $\w$ with respect to
$M_1$ and $M_2$, thereby obtaining an incomplete expression for
$b_1^L$.

The halo correlation function in Lagrangian space, $\xi_{hh}^L$ from
eq.(\ref{eq:k1}), with $M_1=M_2=M$ and $z_1=z_2\equiv z_f$, is shown in
Figure 1 for two scale-free power-spectra, $P(k)\propto k^n$, with
spectral index $n=-2$ and $n=-1$, in the Einstein-de Sitter case. The
two-point function is calculated for various halo masses in units of
the characteristic mass, $M_\ast$, defined so that $\sigma_{M_\ast}
\equiv t_f=\de_c/D(z_f)$, with top-hat filtering\footnote{We are
adopting here the MW definition of $M_\ast$, which, although differing
from the standard PS one, $\sigma_{M_\ast}\equiv t_f/\sqrt{2}$, is more
convenient for our present purposes.}; the spatial dependence is shown
as a function of the scaling variable $q/R$, which eliminates any
residual redshift dependence.  Also shown is the mass autocorrelation
function $\xi_{m}$ and an estimate of the Lagrangian halo two-point
function obtained as $(b_1^L)^2\xi_{m}$, for $M\neq M_\ast$, and
$(b_2^L)^2\xi_{m}^2/2$, for $M=M_\ast$, as in this case $b_1^L=0$. Note
that such an estimate of $\xi_{hh}^L$ always provides an accurate fit
to its analytical expression for separation a few times larger than
the halo size. The characteristic behaviour of the halo correlation
function for $M=M_\ast$, where the linear bias vanishes, is actually a
peculiarity of the Lagrangian case (see also MW). As we will see below,
the Eulerian halo correlation function does not show such a drastic
change of slope in the same mass range.

\begin{figure}
\centering{
\vbox{
\hbox{
\psfig{figure=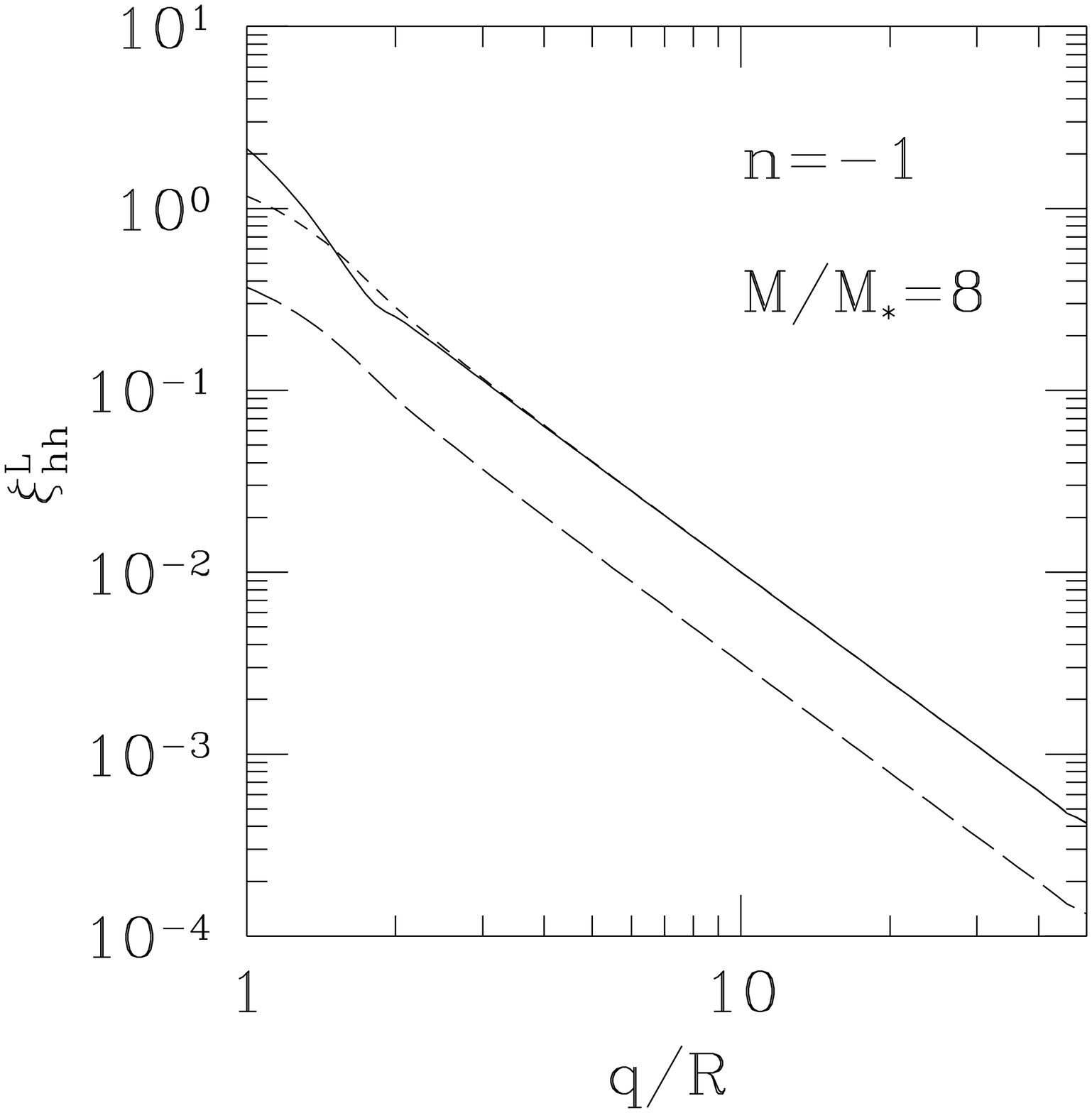,bbllx=20pt,bblly=217pt,bburx=565pt,bbury=700pt,clip=t,width=6.137cm}
\psfig{figure=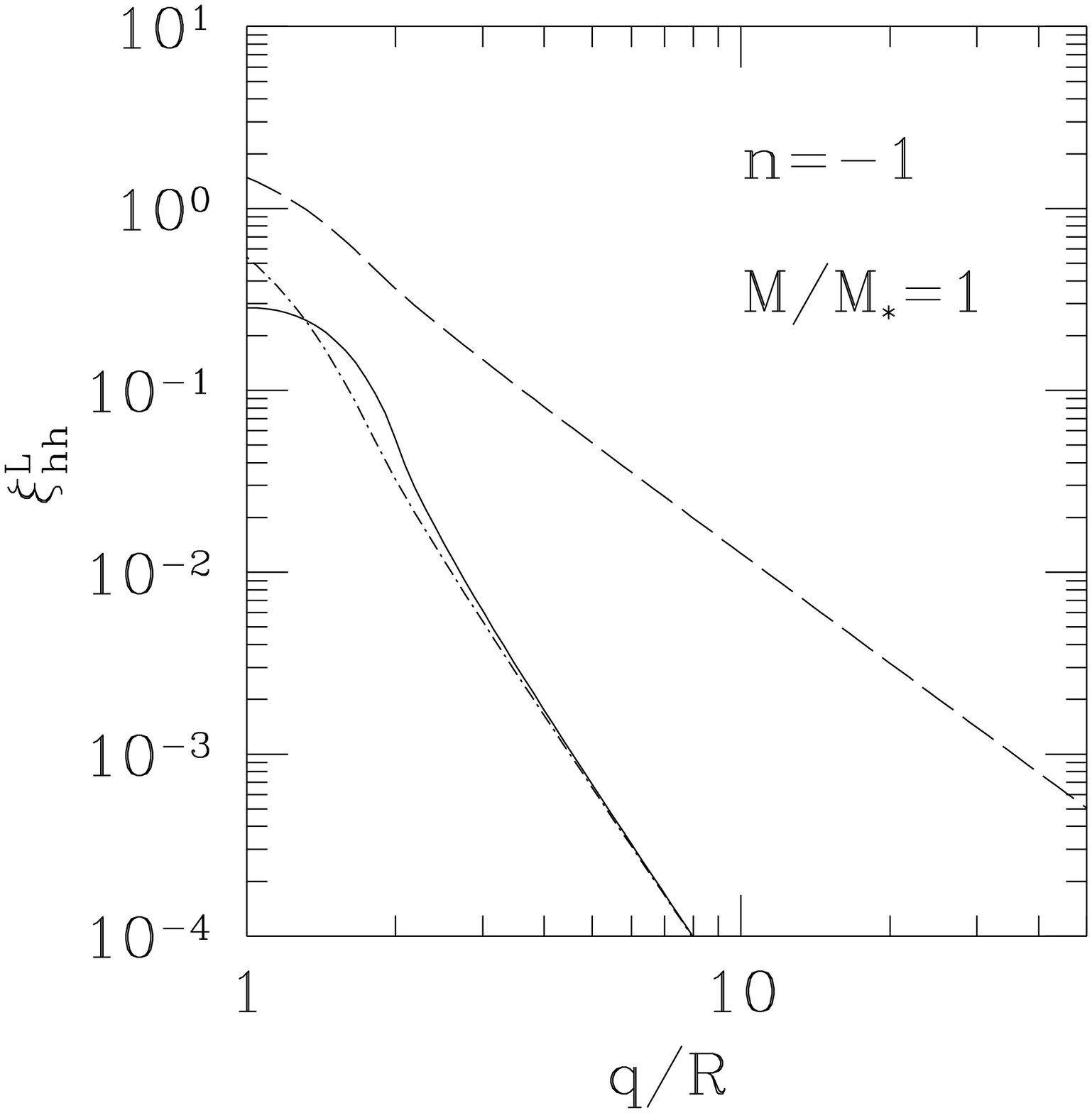,bbllx=139pt,bblly=217pt,bburx=565pt,bbury=700pt,clip=t,width=4.78cm}
\psfig{figure=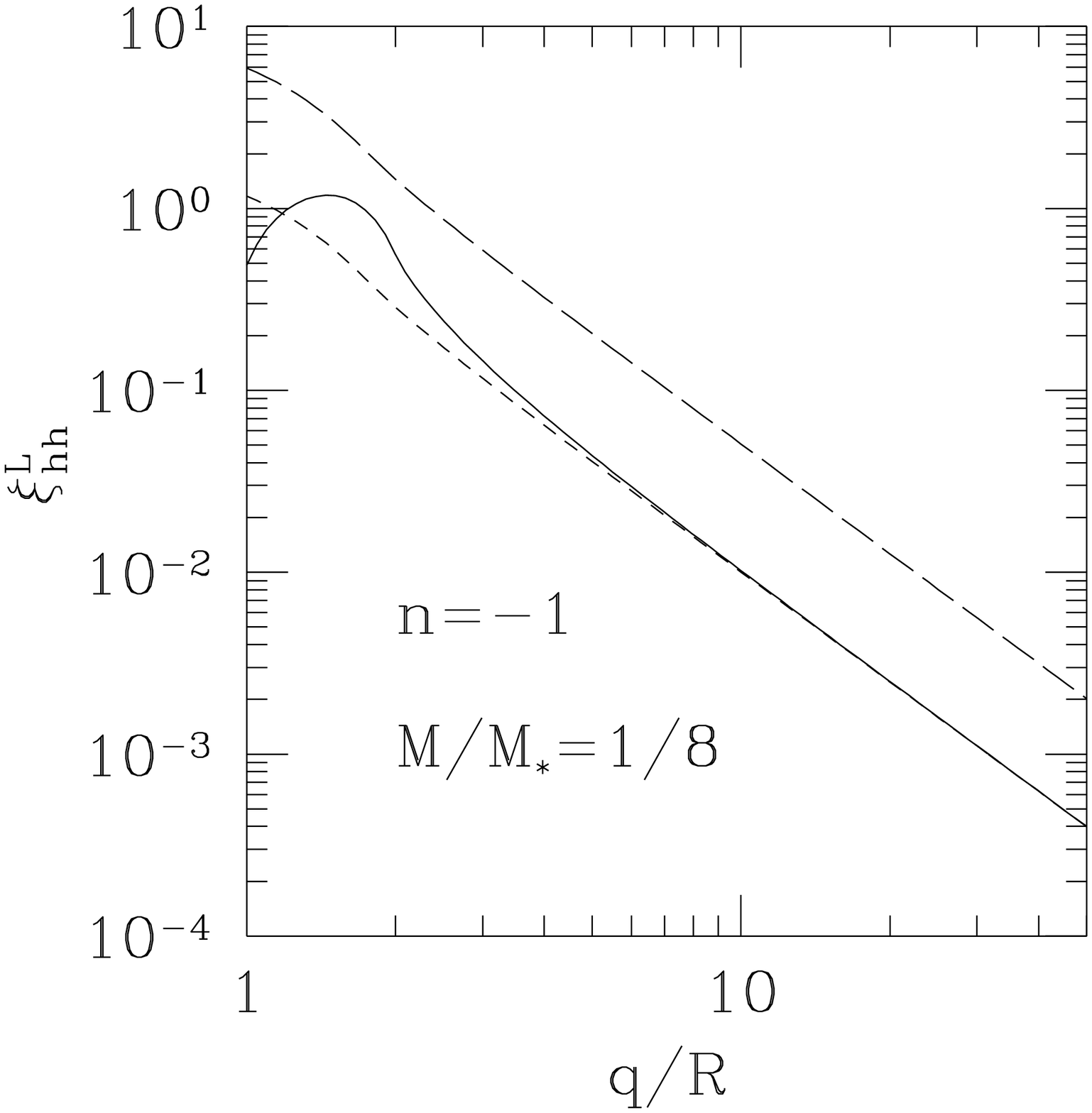,bbllx=139pt,bblly=217pt,bburx=565pt,bbury=700pt,clip=t,width=4.78cm}
}
\hbox{
\psfig{figure=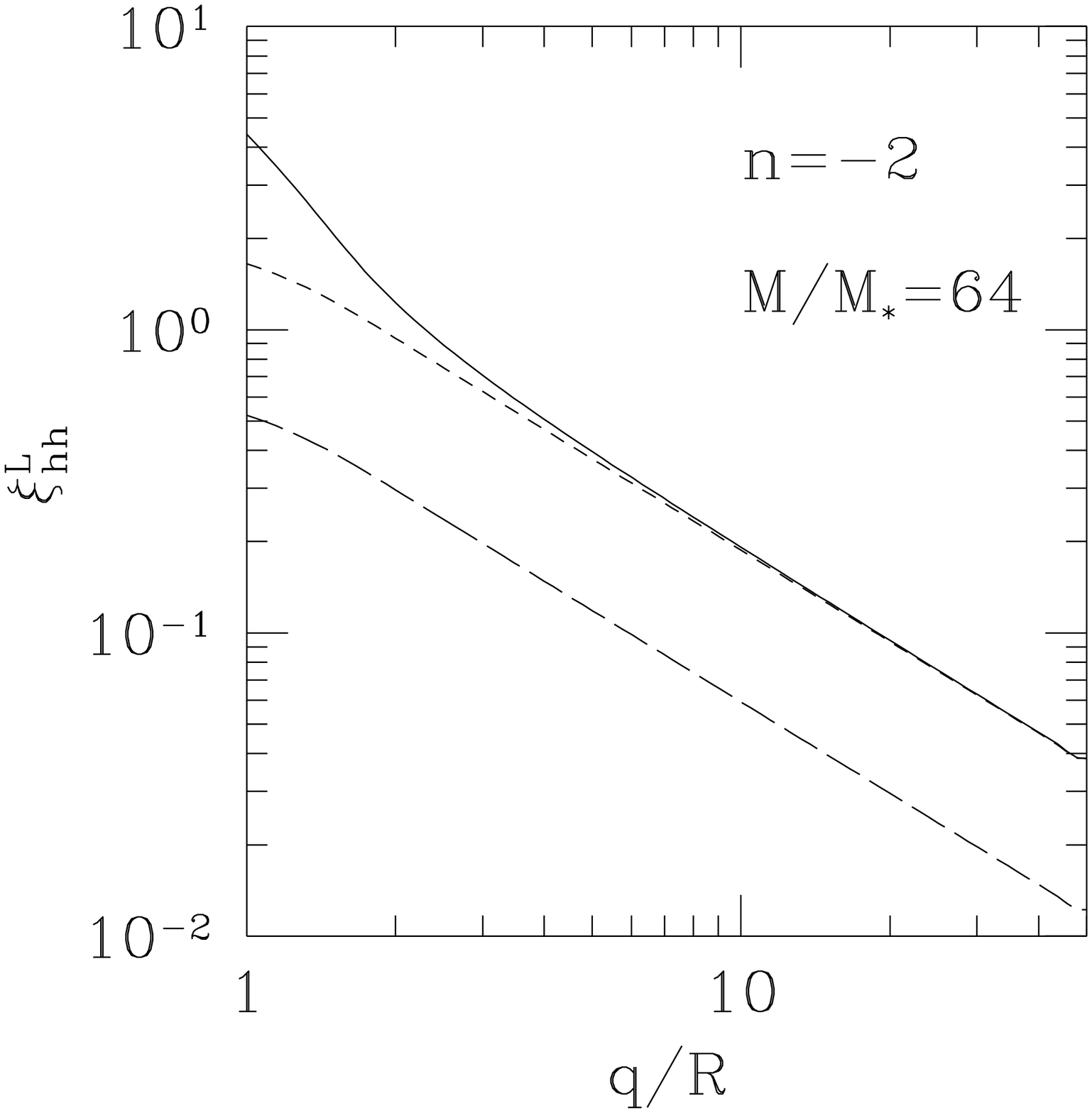,bbllx=20pt,bblly=144pt,bburx=565pt,bbury=700pt,clip=t,width=6.113cm}
\psfig{figure=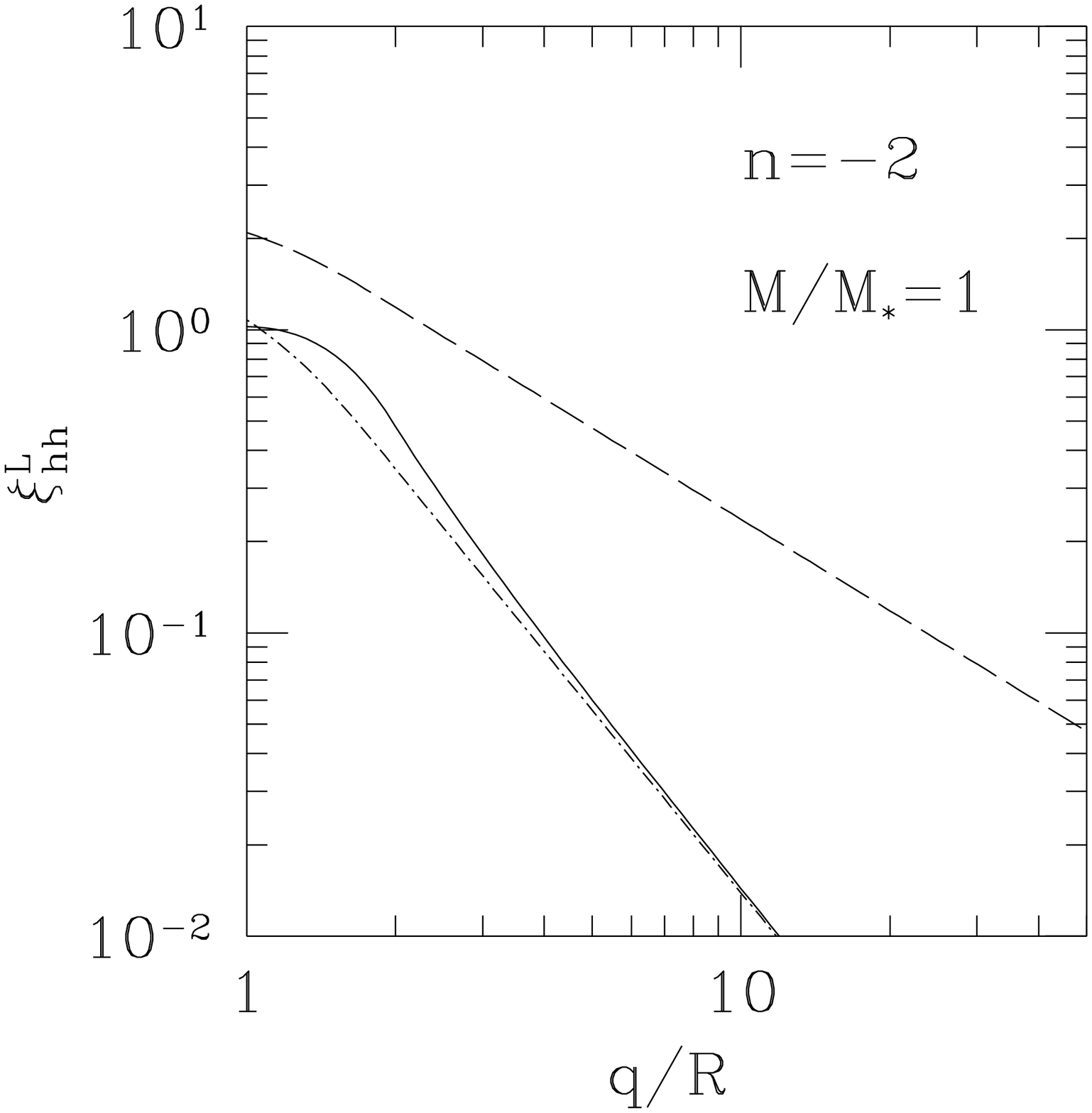,bbllx=139pt,bblly=144pt,bburx=565pt,bbury=700pt,clip=t,width=4.79cm}
\psfig{figure=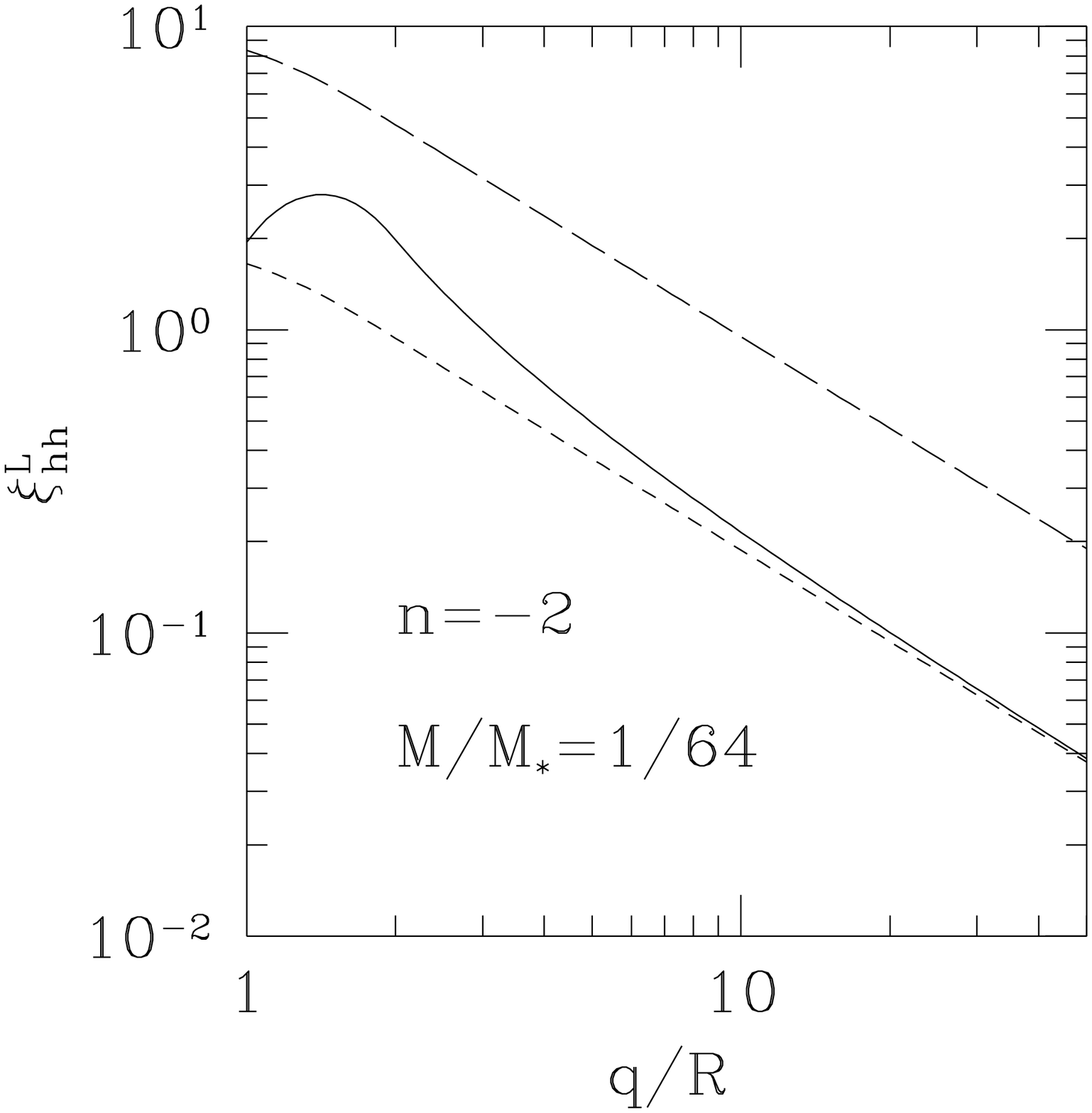,bbllx=139pt,bblly=144pt,bburx=565pt,bbury=700pt,clip=t,width=4.79cm}
}
}
}
\caption{The exact Lagrangian halo correlation function in an
Einstein-de Sitter universe ({\em solid lines}) is shown for scale-free
models with spectral index $n=-1$ and $n=-2$, and for various
masses. In each panel we set $M_1=M_2=M$ and $z_1=z_2$. Results are
plotted in terms of the scaling variables $M/M_\ast$ and $q/R$ (with
$R$ the top-hat radius corresponding to the halo mass $M$), which makes
the resulting curves redshift independent. For comparison, the linear
mass autocorrelation function smoothed on the halo scale is also shown
with {\em long-dashed lines}. The {\em short-dashed lines} represent
the linear bias approximation for the halo correlations: $(b_1^L)^2
\xi_m$.  In the central panels, where $b_1^L=0$, the {\em dot-dashed
lines} show, instead, the second-order approximation for
$\xi_{hh}^L$. Each column contains panels that refer to the same mass
variance ($\sigm^2/t_f^2=1/4,~1,~4$) and so to the same Lagrangian bias
factors. Notice that, for separation a few times the halo size the
first non-vanishing term of eq.(15) always gives an accurate
approximation to the exact halo correlation. This implies that, for
$M_\ast$ objects, $\xi_{hh}^L \approx (b_2^L)^2 \xi_m^2/2$.}
\label{fig:l}
\end{figure}

\subsection{Peak-background split}

In the previous paragraphs we computed number densities and correlation
functions of haloes in Lagrangian space. However, after their
identification, these haloes in embryo move following the gravitational
field, modifying their original spatial distribution. One issue to
address is how, for instance, the conditional halo number density per
unit mass changes as a consequence of gravitational evolution.
Furthermore, of interest is to quantify the evolution of clustering in
terms of the halo correlation functions, or in terms of the
halo-to-mass bias. Both problems can be dealt with by defining Eulerian
halo counting fields, in the same spirit as we did for the Lagrangian
case.

Essentially, our approach to the clustering evolution can be based on a
generalization of the so-called peak-background split, first proposed
by Bardeen \etal (1986), which basically consists in splitting the mass
perturbations in fine-grained ({\em peak}) and coarse-grained ({\em
background}) components\footnote{We are here making a rather liberal
use of the word `peak', to mean the fine-grained component of the
linear density field.}.  The underlying idea is to ascribe the collapse
of objects on small scales to the high-frequency modes of the density
field, while the action of large-scale structures on these non-linear
condensations is due to the remaining low-frequency modes. At the
linear level the resulting effect of these long wavelengths is simply
modeled as a shift of the local background density.

In the spirit of the peak-background split, we define the linear
density field smoothed on a given scale $\eps_M$ as consisting of two
complementary and superimposed components, namely
$\eps_M=\eps_{bg}+\eps_{pk}$.  Adopting as window function the sharp
$k$-space filter, we define as `background' component the density
contrast smoothed on the scale $R_\0=1/k_\0$
\be
\eps_{bg}(\bq, z)\equiv\int \f{d\bk}{(2\pi)^3} \,
\tilde\eps(\bk, z)\, \Theta(k_\0-k)\,  {\rm e}^{i \bk\cdot\bq}\;.
 \ee
The `peak' component is instead obtained by smoothing the mass density
fluctuation with the band-pass filter $\Theta(k_M-k)\,\Theta(k-k_\0)$,
namely
\be
\eps_{pk}(\bq, z)\equiv\int \f{d\bk}{(2\pi)^3}\,
\tilde\eps(\bk, z) \,\Theta(k_M-k)\,\Theta(k-k_\0) \,
{\rm e}^{i \bk\cdot\bq}\;, 
 \ee
where $k_M=1/R_M$, with $M\propto \rho_b R_M^3$ and $M_\0\propto \rho_b
R^3_\0$ the masses enclosed by the two filters.  So, the peak component
contains only modes with wavenumber in the interval $[k_\0,k_M]$.  Note
that in the linear regime, with Gaussian initial conditions, the peak
and background components are statistically independent, i.e.,
\be 
\lan \eps_{pk}(\bq_1, z)\, \eps_{bg}(\bq_2, z) \ran = 0\;, 
 \ee 
for, by construction, the two fields do not share any common Fourier
mode. To summarize: provided the collapsed object is described
according to the spherical model, as in the PS theory, the peak field
$\eps_{pk}(\bq, z)$ can be thought as evolving in a local environment
with effective mean density $\rho_b[1+\eps_{bg}(\bq, z)]$. This implies
that the collapse condition can be written as $\eps_{pk}(\bq, z) =
\de_f(z, z_f) - \eps_{bg}(\bq, z)$.

\subsection{Eulerian halo counting field and bias}

The previous analysis shows how the PS and the conditional Lagrangian
mass functions can be obtained by averaging properly defined halo
counting random fields. It is thus legitimate to explore the
possibility of building up analogous counting processes in the Eulerian
world.  Our approach here will be based on the peak-background split
technique described above.

Let us define the Eulerian counting field of haloes collapsed at 
redshift $z_f$ and observed at $z$ as
\be \calN_h^E(\bq,z|M,z_f) \equiv 
\big[1+\eps_{bg}(\bq,z)\big]\, \calN_h^L(\bq,z|M,z_f) = -\f{2\rho_b}{M} 
\big[1+\eps_{bg}(\bq,z)\big]\, 
\f{\p}{\p M} \Theta\big[\eps_{pk}(\bq, z)
-\big(\de_f(z,z_f)-\eps_{bg}(\bq,z)\big)\big] \;. 
\label{eq:p-b}\ee 
The watchful reader might wonder about our use of the Lagrangian
variable $\bq$ within the Eulerian framework, however, in linear
theory, $\bx=\bq$. Once again, the redshift $z$ must be thought of as
the redshift the sampled objects have at the epoch of observation.
Noteworthy, eq.(\ref{eq:p-b}) is fully consistent with the analysis in
Cole and Kaiser (1989).  Most importantly, our treatment allows for a
local description.  Let us stress here that the factor $(1+\eps_{bg})$,
connecting the Eulerian to the Lagrangian counting field, simply comes
from mass conservation in Eulerian space [see also Section 3.1 and, in
particular, eq.(\ref{eq:intcontin})]; this point has been discussed in
more detail by Kofman \etal (1994).

Now consider the integral stochastic process 
\be
\int_M^\infty d M' M' \calN_h^E(\bq,z|M',z_f)
= 2\,\rho_b\,[1+\eps_{bg}(\bq,z)]\,\Theta [\eps_M(\bq,z) - \de_f(z,z_f)] 
\;;
\label{eq:wbias}
\ee
this represents the fraction of mass, in the unit Eulerian comoving
volume centered in $\bq$, which at redshift $z_f$ will form haloes more
massive than $M$.  For $M_\0 \to M$, $\eps_{bg} \to \eps_M$ and the
above relation coincides (up to the usual fudge factor of $2$, having
no effect on correlations) with the {\em weighted bias} model of
Catelan \etal (1994). An extended version of this scheme, called
`censoring bias', has been recently proposed by Mann, Peacock and
Heavens (1997). Thus, the weighted bias is just the Eulerian version,
within linear theory, of Kaiser (1984) bias model.

Of course, further specifications could be added to our Eulerian
counting field.  For instance, we might ask that the background scale
has not yet collapsed by the epoch $z_f$; in such a case we should
multiply the above stochastic process by the factor $\Theta[\de_f(z,
z_f) - \eps_{bg}(\bq,z)]$. Extra details of this kind would however
make negligible changes to our final results, provided $\sigm^2 \gg
\s_\0^2$.

Like in the Lagrangian case, to calculate the mean halo number density
per unit mass, one needs to ensemble average
$\calN_h^E(\bq,z|M,z_f)$. Let us analyze this operation in more detail.
Because of the way the Eulerian counting process has been defined, it
is clear that $\calN_h^E(\bq,z|M,z_f)$ depends on two random fields,
specifically $\eps_{bg}$ and $\eps_{pk}$.  So, the ensemble average
$\lan\calN_h^E\ran$ can be interpreted as a double average over these
fields, i.e.  $\lan\calN_h^E\ran \equiv
\lan\,\lan\calN_h^E\ran_{\eps_{pk}}\ran_{\eps_{bg}}$. The statistics of
the field $\calN_h^E$ can be described in terms of $n$-th order
correlation functions, $\lan\,\lan\calN_h^E(\bq_1) \cdots
\calN_h^E(\bq_n)\ran_{\eps_{pk}}\ran_{\eps_{bg}}$. The exact
calculation of these quantities is rather difficult. However, because
of the short-scale coherence of the peak field, implied by the
`infrared' cutoff at $k_\0$, its covariance
$\lan\eps_{pk}(\bq_i)\,\eps_{pk}(\bq_i+\br)\ran_{\eps_{pk}}$ vanishes
whenever $r\gg R_\0$, so that we can simplify the general halo
correlations above as $\lan\,\lan\calN_h^E(\bq_1)\cdots
\calN_h^E(\bq_n)\ran_{\eps_{pk}}\ran_{\eps_{bg}} \approx
\lan\,\lan\calN_h^E(\bq_1)\ran_{\eps_{pk}}\cdots
\lan\calN^E(\bq_n)\ran_{\eps_{pk}}\ran_{\eps_{bg}}$, provided we
consider sets of points $\bq_i$, $i=1, \dots, n$, with relative
separation $r_{ij}\equiv |\bq_i - \bq_j| \gg R_\0$.  Therefore, with
the purpose of calculating the mean Eulerian halo number density per
unit mass and Eulerian halo correlations, we can make the replacement
$\calN_h^E \rightarrow \lan\calN_h^E\ran_{\eps_{pk}}\equiv N^E_h$, with
only negligible loss of accuracy. According to the definition of
$\calN^E_h$ in eq.(\ref{eq:p-b}), the latter ensemble average gives
\be
N^E_h(\bq,z|M,z_f)= 
\f{1}{\sqrt{2\pi}}\, \f{\rho_b}{M}
\big[1+\eps_{bg}(\bq, z)\big]
\f{\de_f(z, z_f)-\eps_{bg}(\bq, z)}{[\sigm^2\!(z)-\s^2_\0(z)]^{3/2}}\, 
\exp \left\{-
\f{[\de_f(z, z_f)-\eps_{bg}(\bq, z)]^2}{2\,[\sigm^2\!(z)-\s^2_\0(z)]} 
\right\}\, \left| \f{d\sigm^2\!(z)}{dM} \right|\;, 
\label{eq:cge}
\ee
which, having averaged over the fine-grained mass fluctuations,
represents a sort of coarse-grained halo counting field. Notice that
the fine-grained ensemble average has replaced the original
step-function operator of eq.(\ref{eq:p-b}) by a smoother function,
which can then be consistently expanded in series of the background
field, as shown below.

Let us stress that the expression in eq.(\ref{eq:cge}) is just the
Eulerian analog of eq.(10) in MW, but the field $\eps_{bg}$ is here a
true random field, and so is the process $N^E_h$.  The knowledge of
$N^E_h$ allows us to define the Eulerian halo number density
fluctuation as
\be
\de^E_h(\bq,z|M, z_f)\equiv 
\f{N^E_h(\bq, z|M,z_f)-\lan N^E_h(\bq, z|M,z_f )\ran_{\eps_{bg}}}
{\lan N^E_h(\bq, z|M,z_f )\ran_{\eps_{bg}} } \equiv 
\rmb^E(\bq,z|M,z_f)\,\eps_{bg}(\bq,z)\;, 
\ee
where we introduced the Eulerian `bias field'
$\rmb^E(\bq,z|M,z_f)$. The second equality in the above equation does
not mean that the Eulerian fluctuation field $\de^E_h$ is proportional
to the background density field $\eps_{bg}$. In fact, $\rmb^E$ in
general depends upon $\eps_{bg}$ itself. Its functional dependence can
be understood by expanding $N^E_h(\bq, z|M,z_f )$ in powers of
$\eps_{bg}$ to obtain
\ba
\de^E_h( \bq, z|M,z_f)
& = & b_1^E(z|M,z_f)\,\eps_{bg}(\bq,z) + {1 \over 2} 
\,b_2^E(z|M,z_f) \,\eps_{bg}^2(\bq,z) + \dots \nonumber \\
& = & 
\left[1+b_1^L(z|M,z_f)\right] \eps_{bg}(\bq,z) + {1 \over 2} 
\left[b_2^L(z|M,z_f) + 2\, b_1^L(z|M,z_f)\right] 
\eps_{bg}^2(\bq,z) + \dots \;,  
\label{eq:dehE}
\ea
where, for $\sigm^2 \gg \s_\0^2$, the first and second-order Lagrangian
bias factors $b_1^L$ and $b_2^L$ are those of eq.(\ref{eq:b1}) and
eq.(\ref{eq:b2}), respectively. Accounting for the transformation from
the Lagrangian to the Eulerian distribution (e. g. Kofmann \etal 1992),
one has $\lan N^E_h(\bq, z|M,z_f)\ran_{\eps_{bg}} = \nps(z|M,z_f)$.  It
can be useful to give explicit expressions for the first two Eulerian
bias parameters of linear theory
\be
b^E_1(z|M,z_f) = 1 + \f{D(z_f)}{D(z)}
\left[\f{\de_c}{D(z_f)^2\sigm^2}-\f{1}{\de_c}\right]\;,
\label{eq:bhE}
\ee
\be
b^E_2(z|M,z_f) = \f{1}{D(z)^2\sigm^2} \left[ \f{\de_c^2}{D(z_f)^2\sigm^2}-3 
\right] + \f{2 D(z_f)}{D(z)} \left[\f{\de_c}{D(z_f)^2\sigm^2} - 
\f{1}{\de_c}\right] \;.  
\label{eq:bhE2}
\ee

The set of linear theory Eulerian bias factors $b^E_\ell(z)$ can be
obtained from the Lagrangian ones according to the general rule
\be 
b^E_\ell = \ell\,b^L_{\ell-1} + b^L_\ell \;, 
\label{eq:itera}
\ee 
with $b^L_{\ell=0} \equiv 1$. 

The same method can be applied to the Lagrangian expression, in the
sense that we can obtain, similarly, 
\be 
N^L_h(\bq,z|M,z_f)=
\f{1}{\sqrt{2\pi}}\, \f{\rho_b}{M} \f{\de_f(z, z_f)-\eps_{bg}(\bq,
z)}{[\sigm^2\!(z)-\s^2_\0(z)]^{3/2}}\, \exp \left\{- \f{[\de_f(z,
z_f)-\eps_{bg}(\bq, z)]^2}{2\,[\sigm^2\!(z)-\s^2_\0(z)]} \right\}\,
\left| \f{d\sigm^2\!(z)}{dM} \right|\;.
\label{eq:cgl}
\ee
One has, exactly, $\lan N^L_h(\bq, z|M,z_f)\ran_{\eps_{bg}} = \lan
{\cal N}^L_h(\bq, z|M,z_f)\ran = \nps(z|M,z_f)$.  By expanding the
coarse-grained Lagrangian counting field $N^L_h(\bq,z|M,z_f)$ we can
define Lagrangian bias factors at any order. For $\sigm^2 \gg \s_\0^2$
these turn out to be identical to those deriving from the expansion of
the halo correlation in Lagrangian space, eq.(\ref{eq:k1}). This
suggests, however, that these bias factors can be used to describe halo
clustering on distances $r > R$, without any further restriction
introduced by the background scale $R_\0$.

The very fact that, for practical purposes, we can replace the exact
operator $\calN^E_h$ by the locally averaged one $N^E_h$ demonstrates
that the MW treatment can be made self-consistent, provided their
small-scale density field is replaced by the peak field, and the value
of the threshold is modified accordingly. Most importantly, our local
averaging procedure implies that, up to the scale $R_\0$, we are indeed
correctly accounting for the cloud-in-cloud problem. This is because at
each point $\bq$, characterized by a random value of the background
field $\eps_{bg}(\bq)$, the coarse-grained stochastic process
$N^E_h(\bq, z|M,z_f )$ (and its Lagrangian equivalent) actually
represents the local mean mass function, for which the cloud-in-cloud
problem is exactly solved in terms of first passage `times' across the
local barrier $\delta_f(z,z_f) - \eps_{bg}(\bq ,z)$, with initial
condition $\eps_{pk}(\bq,z)=0$ at $R=R_\0$. Therefore, to the aim of
calculating correlations on lags $r \gg R_\0$, we can safely state that
our coarse-grained halo counting fields are unaffected by the
cloud-in-cloud problem.

The shift by 1 of the linear bias factor, here implied by the
transformation from the Lagrangian to the Eulerian world, was also
noticed in the weighted bias approach by Catelan \etal [1994; their
eq.(21)], where an underlying Lognormal distribution was assumed to
avoid negative-mass events.

The above expression for $b^E_1(z|M,z_f)$ coincides with the formula by
MW [their eq.(20)], who, however, only presented it for $z=0$.  As
noticed by MW, an important feature of this linear bias is that it
predicts that large-mass objects (actually those characterized by
$\sigm < t_f$) are biased with respect to the mass ($b_1^E>1$), while
small-mass ones ($\sigm > t_f$) are anti-biased ($b_1^E<1$).  Haloes
with mass close to the characteristic one, $M_\ast$, have non-vanishing
linear bias, unlike the Lagrangian case. As we will see in Section 3.1,
this one-to-one classification of biased and anti-biased objects
according to their mass is no longer valid in the non-linear regime, as
the shear field at the Lagrangian location of the halo also contributes
to the determination of its Eulerian bias factor.

The effect of merging can be easily accommodated into this scheme.  In
the real Universe, haloes undergo merging at some finite rate, which
can be suitably modeled (e.g. Lacey \& Cole 1993). As mentioned above,
in the simple PS theory such a rate is actually infinite, for infinite
mass resolution, implying that only haloes `just formed' can survive,
so that $z_f=z$. So, if one gives up singling out the individuality of
haloes selected at different threshold, i.e. with different formation
redshifts $z_f\geq z$, one immediately obtains (e.g. Matarrese \etal
1997)
\be 
b_1^E (z|M)= 1+ \left[{\de_c \over D(z)^2\sigm^2}- {1 \over
\de_c}\right] \;, 
\ee 
which implies a quadratic redshift dependence in the
Einstein-de Sitter universe,
\be b_1^E (z|M)= 1+ \left[{\de_c (1+z)^2 \over \sigm^2}- {1 \over
\de_c}\right] \;.
\ee
The latter form coincides with the result by Cole and Kaiser (1989)
[their eq.(6)], who however define the bias factor of haloes at
redshift $z$ with respect to the mass fluctuation at the present time,
which then scales the latter expression by a factor $(1+z)^{-1}$.

On the other hand, for fixed $z_f$ and varying $z$, i.e. for objects
which survived till the epoch $z$ after their birth at $z_f$, the 
Eulerian bias of eq.(\ref{eq:bhE}) gets a completely different evolution, 
namely
\be 
b_1^E (z|M)= 1+ \f{D(z_f)}{D(z)} \Big[b_1^E (z_f|M) -1 \Big] \;,
\ee
which implies a linear redshift dependence in the Einstein-de
Sitter case,
\be
b_1^E (z|M)= 1+ \f{1+z}{1+z_f} \Big[b_1^E (z_f|M) -1 \Big] \;.
\ee
The latter form coincides with that obtained by Dekel (1986), Dekel and
Rees (1987), Nusser and Davis (1994) and Fry (1996). This relation can
be relevant for galaxies which were conserved in number after their
formation, i.e. that maintained their individuality even after their
hosting haloes merged.

It is trivial, at this point, to obtain the Eulerian halo-halo
correlation function within our approximations. For lags $r \gg R_\0$,
one has
\be 
\xi_{hh}^E(r, z|M_1,z_1; M_2,z_2) = b^E_1(z|M_1,z_1)\, b^E_1(z|M_2,z_2)\,
\xi_m(r,z) + \f{1}{2}\,b_2^E(z|M_1,z_1)\,b_2^E(z|M_2,z_2)\,\xi^2_m(r,z) + 
\cdots\;. 
\ee
The main limitation of this formula, however, is that it only provides
a link between the Eulerian halo correlation function and that of the
mass within linear theory. What one would really need, instead, is a
similar relation in the fully non-linear regime. This problem will be
solved in the next section.

\section{Halo counting and non-linear dynamics: Eulerian description}

One can derive a general expression for the Eulerian halo-to-mass bias
by integrating the continuity equations for the mass and for the halo
number density, assuming that haloes move according to the velocity
field determined by the matter.  The Lagrangian analysis carried out in
the previous section is crucial to the present purposes, since it
allows for the natural initial conditions necessary to integrate the
Eulerian equations. As we will show below, the Eulerian halo-to-mass
bias obtained in such a way holds for any cosmology and in any
dynamical regime. This turns out to be a remarkable generalization of
the biasing proposed by Cole and Kaiser (1989) and MW.

\subsection{Eulerian bias from dynamical fluid equations}

Let us consider the mass density fluctuation field $\de(\bx,
\tau(z))=\de(\bx, z)$ which obeys the mass conservation equation
\be
\f{d\de}{d \tau}=-(1+\de)\nabla\cdot\bfv \;,
\ee
where $\tau$ is the conformal time of the background cosmology and the
differential operator $d/d\tau\equiv \p/\p \tau + \bfv\cdot\nabla$ is
the convective derivative. The peculiar velocity field $\bfv\equiv
d\bx/d \tau$ satisfies the Euler equation $ d\bfv/d\tau + (a'/a)\bfv =
- \nabla \phi_g$, where $a$ is the expansion factor and a prime denotes
differentiation with respect to $\tau$. For later convenience, let us
also define the scaled peculiar velocity $\bfu \equiv d\bx/d D = \bfv /
D'$.  The peculiar gravitational potential $\phi_g$ is determined by
the matter distribution via the cosmological Poisson equation
$\nabla^2\phi_g = 4\pi G a^2 \rho_b(\tau) \de$, where $\rho_b(\tau)$ is
the background mean density at time $\tau$.  If we assume that our halo
population of mass $M$ and formation redshift $z_f$ is conserved in
time, and evolves exclusively under the influence of gravity, its
number density fluctuation $\de_h(\bx,z) = \de_h(\bx,z|M,z_f)$ has to
satisfy the continuity equation (e.g. Fry 1996)
\be
\f{d\de_h}{d\tau}=-(1+\de_h)\nabla\cdot\bfv\;,
 \ee
from which, eliminating the expansion scalar $\nabla\cdot\bfv$, we
obtain
\be
\f{d\ln(1+\de_h)}{d\tau}=\f{d\ln(1+\de)}{d\tau}\;.
\ee
This equation can be integrated exactly in terms of Lagrangian
quantities, and the solution reads
\be
1+\de_h(\bx, z)=\big[1+\de_h(\bq)\big]\big[1+\de(\bx, z)\big]
\label{eq:intcontin}\ee
(see also the discussion in Peacock \& Dodds 1994), where $\bq$ is the
Lagrangian position corresponding to the Eulerian one via
$\bx(\bq,z)=\bq + \bS(\bq,z)$, with $\bS(\bq,z)$ the displacement
vector.  In eq.(\ref{eq:intcontin}), by $\de_h(\bq) = \de_h(\bq|M,z_f)$
we mean the Lagrangian halo density fluctuation, whereas, for
simplicity, we assumed that $\lim_{z\rightarrow \cal1}\de[\bx(\bq, z),
z]\equiv\de(\bq)=0$, i.e. that the mass was initially uniformly
distributed (this amounts to taking purely growing-mode initial
perturbations). Defining the Eulerian halo bias field through
\be
\de_h(\bx, z) \equiv \rmb^E(\bx, z)\,\de(\bx, z)\;,
\ee
we end up with the {\em exact} relation
\be
\rmb^E(\bx, z)=1+\f{1+\de(\bx, z)}{\de(\bx, z)}\,\de_h(\bq)\;.
\ee
The key problem now is how to calculate the field $\de_h(\bq)$. We
cannot simply take the Lagrangian halo distribution as
$\de_h(\bq)=\rmb^L(\bq)\de(\bq)$, because $\de(\bq)=0$; thus, we are
forced to adopt some limiting procedure. To specify the Lagrangian halo
distribution, we can take advantage of the results of Section 2.  By
definition, the Lagrangian distribution of nascent haloes of mass $M$
and formation epoch $z_f$ is given by
\be
\de_h(\bq|M,z_f) \equiv 
\lim_{z\rightarrow\cal1}\,\rmb^E(\bq,z|M,z_f)\,\eps_{bg}(\bq,z)
\equiv \rmb^L_\0 (\bq|M,z_f)\,\eps_\0(\bq) \;, 
\ee
where $\rmb_\0^L(\bq|M,z_f) \equiv \rmb^L(\bq,z\!=\!0|M, z_f)$ is the
Lagrangian halo bias field. Once again, let us stress that the second
equality in the latter equation does not mean at all that $\de_h(\bq)$
is proportional to $\eps_\0(\bq)$. In fact, $\rmb^L$ is in general a
functional of the background density field.  To understand the above
equation, one has to remember that, at sufficiently early times, the
expression for the Eulerian bias field obtained in linear theory
becomes exact (as linear theory gets more and more accurate), and
$\de(\bx,z)\rightarrow \eps_{bg}(\bx, z)=D(z)\,\eps_\0(\bq)$, as
$z\rightarrow \cal1$. Because of our normalization of $D$, here
$\eps_\0(\bq)$ is the mass density fluctuation linearly extrapolated to
the present time and filtered on the background scale $R_\0$.  The
background smoothing scale $R_\0$ actually has a twofold role in our
analysis.  In the linear theory approach of Section 2 it was introduced
and required to be much larger than the halo size, in order to get a
self-consistent definition of halo counting fields, with the desirable
feature of being free of the cloud-in-cloud problem. In the present
non-linear analysis, however, the background mass scale must be chosen
large enough to ensure that the halo velocity field coincides with the
one of the matter.

The Lagrangian density contrast of haloes identified by a PS-type
algorithm can be obtained from eq.(\ref{eq:cgl}) as 
$\de_h(\bq|M,z_f) = N^L_h(\bq, z|M,z_f)/\nps(z|M,z_f) - 1$, 
which leads to
\be
\de_h(\bq|M,z_f) = \left[1 -  \f{D(z_f) \eps_\0(\bq)}{\de_c} 
\right] \left(1 -  \f{\s_\0^2}{\sigm^2} \right)^{-3/2}  
\exp\left[ - \f{\eps_\0(\bq)^2 - 2 \eps_\0(\bq) \de_c/D(z_f)
+\de_c^2\s_\0^2/D(z_f)^2\sigm^2}{2(\sigm^2-\s_\0^2)} \right] - 1 \;.
\label{eq:exactlb}
\ee
For $\sigm^2 \gg \s_\0^2$ this expression simplifies to 
\be
\de_h(\bq|M,z_f) = 
 \left[1 -  \f{D(z_f) \eps_\0(\bq)}{\de_c} \right]   
\exp\left[ - \f{\eps_\0(\bq)^2 - 2 \eps_\0(\bq) \de_c/D(z_f)}
{2\sigm^2} \right] - 1 =
\sum_{\ell=1}^\infty \f{b_{\0 \ell}^L(M,z_f)}{\ell!} 
\,\eps_\0(\bq)^\ell \;. 
\label{eq:approxlb}
\ee
The first four Lagrangian bias factors evaluated at $z=0$ read 
\be
b_{\0 1}^L(M,z_f) = D(z_f) 
\left[\f{\de_c}{D(z_f)^2\sigm^2}-\f{1}{\de_c}\right]\;,
\label{eq:firstlb} 
\ee
\be
b_{\0 2}^L(M,z_f) = \f{1}{\sigm^2}  \left[ 
\f{\de_c^2}{D(z_f)^2\sigm^2}-3 \right]\;, 
\ee
\be
b_{\0 3}^L(M,z_f) = \f{D(z_f)}{\sigm^2} \left[\f{\de_c^3}{D(z_f)^4 \sigm^4}
- \f{6~\de_c}{D(z_f)^2\sigm^2} + \f{3}{\de_c} \right] \;, 
\ee
\be
b_{\0 4}^L(M,z_f) = \f{1}{\sigm^4} \left[\f{\de_c^4}{D(z_f)^4 \sigm^4}
- \f{10~\de_c^2}{D(z_f)^2\sigm^2} + 15 \right] \;. 
\ee
Note that, in full generality, $b_{\0 \ell}^L(M,z_f)= D(z)^\ell
~b_{\ell}^L(z|M,z_f)$.  Adding the further requirement that the local
fluctuation on the background scale $R_\0$ has not collapsed yet by the
time of halo formation would make our object number density
semi-positive definite both at the Lagrangian and Eulerian level,
i.e. $\de_h \geq -1$, at any time, only provided $\eps_\0 \leq t_f$.

The general expression for the Lagrangian halo density contrast of
eq.(\ref{eq:exactlb}) is plotted in Figure 2 as a function of the
background density field, for different halo masses. In the high-mass
case positive mass fluctuations typically correspond to positive values
of the Lagrangian halo density contrast (and viceversa), while the
trend is the opposite at low masses.  The transition, once again,
corresponds to halo masses around $M_\ast$, in which case positive
values of $\de_h$ only occur in regions with background density close
to the mean.  Also shown are two approximations to the Lagrangian halo
density contrast obtained by expanding eq.(\ref{eq:exactlb}) up to
first and second order in the background field. Except for halo masses
near $M_\ast$, where a quadratic bias is clearly needed, a linear
Lagrangian bias generally provides an accurate fit to $\de_h(\bq)$
within the bulk of the $\eps_\0$ distribution.

\begin{figure}
\centerline{\epsfxsize= 5.6 cm \epsfbox{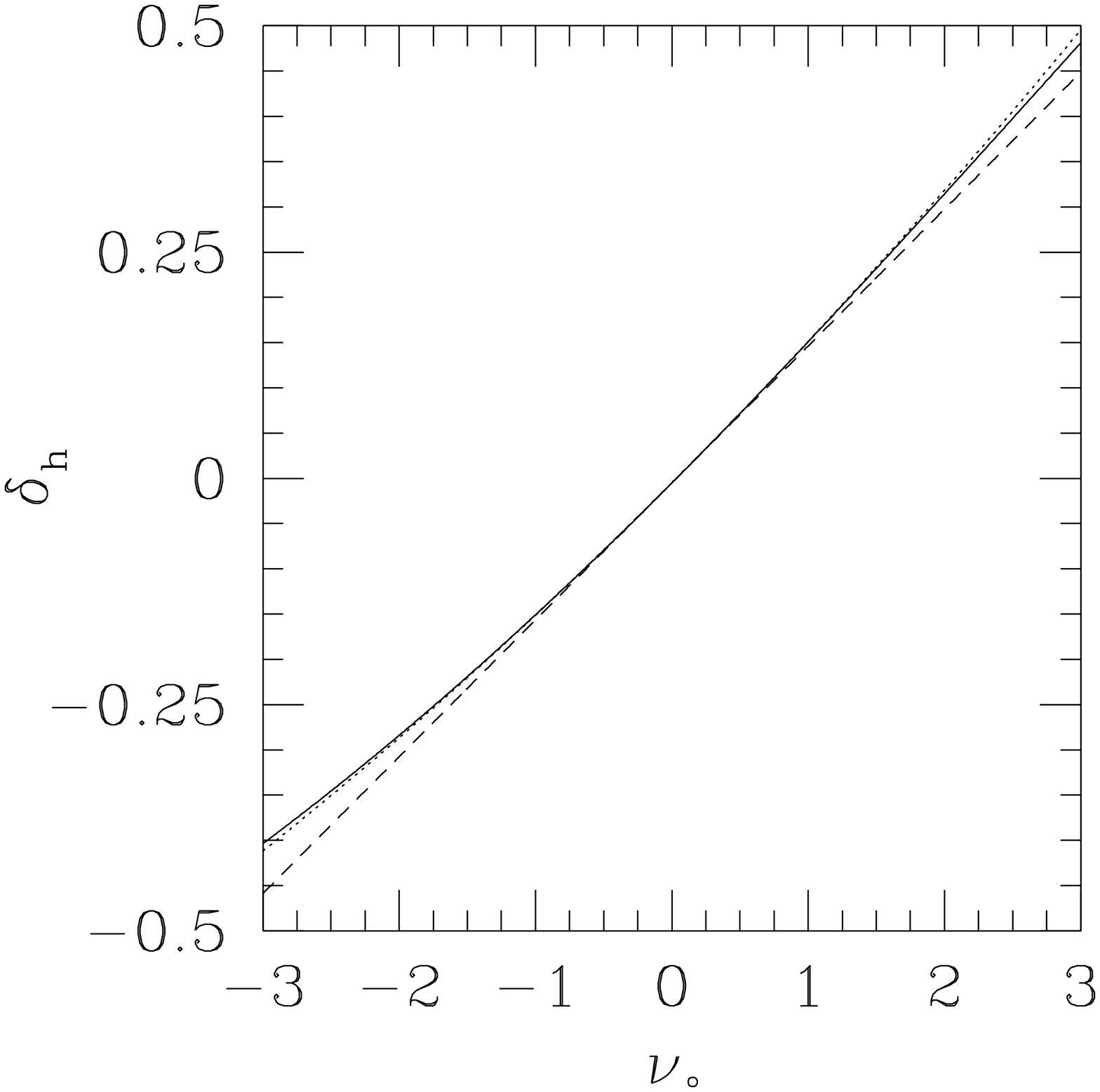}\epsfxsize= 5.6 cm
\epsfbox{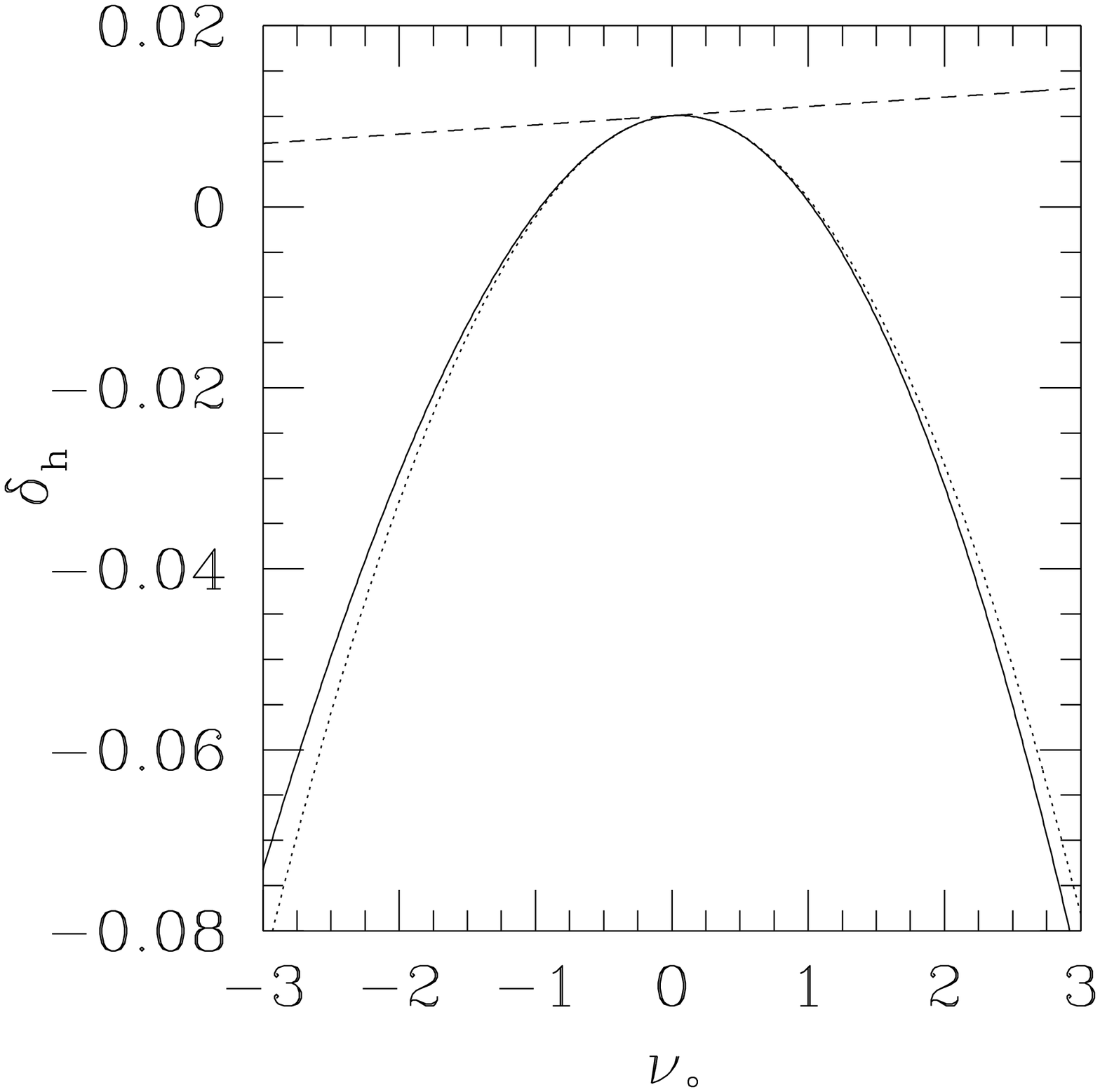}\epsfxsize= 5.6 cm \epsfbox{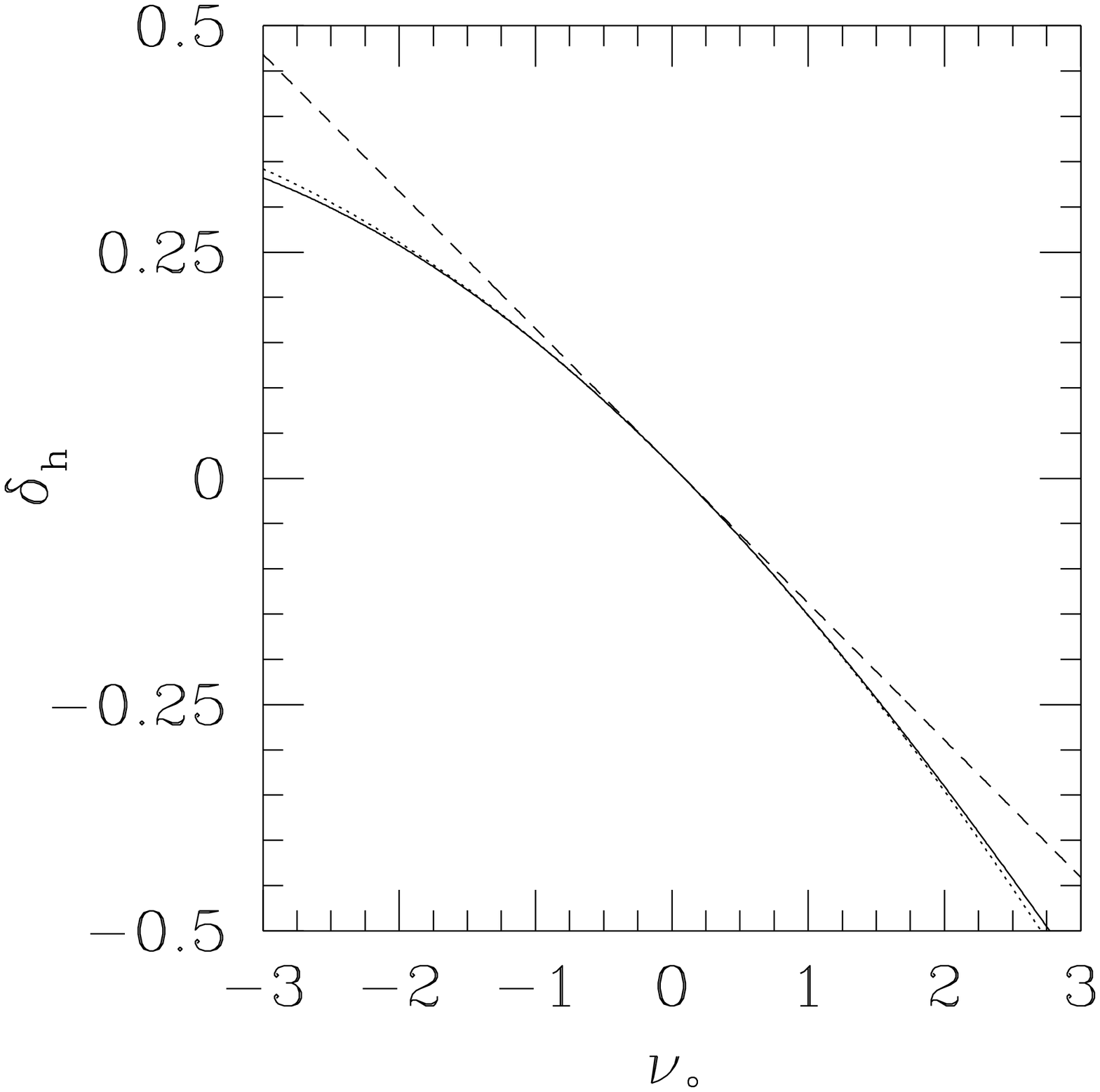}}
\caption{The exact expression for the Lagrangian halo density contrast
of eq.(42) ({\em solid lines}) is plotted as a function of
$\nu_\0\equiv \eps_\0/\s_\0$. The three panels refer to values of the
halo masses such that $\sigm^2/t_f^2=1/4$ ({\em left}),
$\sigm^2/t_f^2=1$ ({\em centre}) and $\sigm^2/t_f^2=4$ ({\em right}),
with $t_f=1.69/D(z_f)$. The background mass scale is chosen so that
$\s_\0^2 = 0.01\,\sigm^2$. Also plotted are two estimates of $\de_h$
obtained by expanding the RHS of eq.(42) up to first ({\em dashed
lines}) and second order ({\em dotted lines}) in $\eps_\0$. Note that,
because of our choice of variables, all the curves are independent both
of $z_f$ and $\Om_\0$.}
\label{fig:2}
\end{figure}

The Eulerian bias field finally reads
\be 
\rmb^E(\bx , z|M, z_f) = 1 + \f{1+\de(\bx , z)}
{\de(\bx , z)}\,\rmb_\0^L(\bq|M,z_f)\,\eps_\0(\bq)\,\;.
\label{eq:biasris}
\ee
It can be seen that, in the linear regime, where $\de(\bx,z)\approx
D(z)\,\eps_\0(\bq) \ll 1$, the expression for $\rmb^E$ in MW [i.e. our
eq.(\ref{eq:bhE})] is recovered, provided $\rmb_\0^L(\bq|M,z_f)$ is
replaced by its first-order approximation $b_{\0 1}^L(M,z_f)$.  It is
however important to realize that the exact expression in
eq.(\ref{eq:biasris}) implies that the Eulerian bias field of dark
matter haloes $\rmb^E(\bx,z|M, z_f)$ is both non-linear, in that it
depends on $\de(\bx)$, and non-local, as it depends on the Lagrangian
position $\bq$ through $\rmb_\0^L(\bq|M,z_f)\,\eps_\0(\bq)$, simply
because of inertia.

Our exact expression for the Eulerian halo bias, eq.(\ref{eq:biasris}),
generally involves quantitative corrections to the MW approximate bias
formula. In some cases, however, the MW relation may even fail to
predict the correct qualitative behaviour of the halo-to-mass bias.
This is the case, in fact, of those initially underdense fluid elements
in Lagrangian space, $\eps_\0(\bq) < 0$, which, after an initial
expansion phase, turn around to undergo a phase of local compression,
so that the corresponding Eulerian fluid element eventually becomes
overdense, $\de(\bx(\bq,z),z) > 0$, and collapses.  This is a
well-known non-linear effect caused by the shear component of the
velocity field, i.e. by the tidal force of the surrounding matter. For
Gaussian initial conditions, the occurrence of such an event can be
estimated by the Zel'dovich approximation as affecting $42\%$ of the
overall Lagrangian volume (Doroshkevich 1970; Shandarin \& Zel'dovich
1984); Hui and Bertschinger (1996), using a different approximation,
estimated this effect as affecting at least $39\%$ of the total
Lagrangian volume.  In all such cases the MW formula would incorrectly
predict bias instead of anti-bias, for halo masses $M > M_\ast$, and
anti-bias instead of bias, for $M < M_\ast$. The problem may be
generally less severe than the above heuristic argument would suggest,
as, at a fixed epoch $z$, only a smaller fraction of such Lagrangian
patches have already turned around from their initial expansion; this
is even more true for large mass haloes, which probe the underlying
mass distribution in a more linear regime, where the MW formula gets
closer to the exact one. As a tentative conclusion let us say that one
should be careful in applying the linear MW bias [i.e. our eq.(\ref{eq:bhE})]
at the Eulerian level especially in connection with halo masses much smaller 
than $M_\ast$.

The most important application of eq.(\ref{eq:biasris}) is that it
allows to generate Eulerian maps of the local comoving halo number
density per unit mass, $\nps(M,z_f)[1 + \de_h(\bx, z|M,z_f)]$, given
the non-linearly evolved mass density contrast $\de(\bx, z)$ (with
Lagrangian resolution $R_\0$) and the corresponding Lagrangian mass and
halo density fluctuation fields, $\eps_\0(\bf q)$ and
$\de_h(\bq|M,z_f)$, respectively.

In order to account for halo merging, at this level, one just has to
assume a suitable link between the formation and observation epochs,
which, in the simple PS theory amounts to the replacement $z_f \to z$,
in the above expressions for $\rmb_\0^L$.

Recalling that mass conservation can be recast in terms of the Jacobian
determinant $J\equiv |\!| \p \bx/\p \bq |\!|$ of the mapping $\bq
\rightarrow \bx$, as $1+\de[\bx(\bq, z), z]=J(\bq, z)^{-1}$, one finds
the exact relation
\be
\rmb^E\left(\bx(\bq, z), z|M, z_f\right) = 1 + 
\left[1-J(\bq, z)\right]^{-1}\,\rmb_\0^L(\bq|M,z_f)\,\eps_\0(\bq) \;.
\label{eq:biasjac}
\ee
It can be useful to illustrate the meaning of this expression by
considering various approximations to the evolution of the mass density
in the non-linear regime, i.e. to the particle trajectories
$\bx(\bq,z)$.  Such approximation schemes should be thought of, not as
self-consistent perturbative approaches to the actual dynamics, but as
`clever tricks' able to catch some aspects of the true dynamics, at
least in the mildly non-linear regime.  A detailed and systematic
comparison of the performance of several approximations for different
choices of the initial conditions has been made by Sathyaprakash \etal
(1995).

\subsubsection{Zel'dovich approximation}

In the Zel'dovich approximation (ZEL; Zel'dovich 1970) the displacement
vector is $\bS = - D \, \nabla_\bfq \varphi_\0(\bq)$, where
$\varphi(\bq)$ is the linear peculiar gravitational potential, suitably
rescaled so that $\nabla^2_\bfq \varphi_\0(\bq) = \eps_\0(\bq)$.
Indicating by $\lam_\al(\bq)$ ($\al=1,2,3$) the eigenvalues of the
deformation tensor $\p^2\varphi_\0(\bq)/\p q_\al\,\p q_\beta$, we
obtain for the Eulerian bias field
\be
\rmb^E_{ZEL}(\bx(\bq,z), z|M,z_f) =
1 + \f{\eps_\0(\bq)\,\rmb_\0^L(\bq|M,z_f)}
{1-\prod_{\al=1}^3\big[1-D(z)\,\lam_\al(\bq)\big]}
= 1 + \f{\rmb_\0^L(\bq|M,z_f)}{D(z)}
\Big[1-D(z)\f{\mu_2(\bq)}{\mu_1(\bq)}+
D(z)^2\f{\mu_3(\bq)}{\mu_1(\bq)}\Big]^{-1}\;. 
\label{eq:zel}
\ee
Here $\mu_1(\bq)\equiv \lam_1+\lam_2+\lam_3 = \eps_\0(\bq)$, $\mu_2
\equiv \lam_1\lam_2+\lam_1\lam_3+\lam_2\lam_3$ and $\mu_3 \equiv
\lam_1\lam_2\lam_3$ are the three invariants of the deformation tensor.
If one makes the further approximation of replacing the Lagrangian bias
by its first-order estimate of eq.(\ref{eq:firstlb}), it can be checked
that the expression of $\rmb^E_{ZEL}$ coincides with the MW result,
both at sufficiently early times ($D\ll 1$) and in the case of
one-dimensional perturbations, for which $\mu_2=0=\mu_3$ and the
Zel'dovich approximation represents the exact solution to the
non-linear dynamics.

It is important to stress that we are not forced to take the above
result as a perturbative expression.  An accurate approximation to the
Eulerian bias field would in fact consist in evolving the mass
according to the truncated (on the scale $M_\0$) Zel'dovich
approximation (Kofman 1991; Kofman \etal 1992; Coles, Melott \&
Shandarin 1993) and using the full expression for the Lagrangian bias.
Being a random field, the Eulerian halo bias is completely
characterized by a probability density functional, thus for a given
mass $M$ and formation redshift $z_f$ there exists a whole distribution
of possible values of $\rmb^E$, related to the particular environment
where the object forms as well as to the initial conditions leading to
that site.  Starting from the ZEL expression in eq.(\ref{eq:zel}) one
could explicitly obtain the probability distribution function
$p(\rmb^E_{ZEL})\,d\,\rmb^E_{ZEL}$, by integrating over the joint
distribution of the invariants $\mu_\alpha$ [an expression for the
latter is given in Kofman \etal (1994)].  These specific applications
of our results will be discussed elsewhere.

Equation (\ref{eq:zel}) has the merit of clearly displaying the
intrinsic non-locality of the Eulerian bias. Only in some simplified
cases there exists a local mapping between $\rmb^E$ and $\de$, so that
an expansion of the halo density contrast in a hierarchy of Eulerian
bias factors, $b^E_1$, $b^E_2$, etc., makes sense. One example is
provided by the linear theory approach of Section 2.6; further examples
are given below.

\subsubsection{Frozen-flow approximation}

According to the {\em frozen-flow} approximation (FFA; Matarrese \etal
1992) the Eulerian density field can be written as
\be
1+\de(\bx(\bq,z), z)=
\exp\,\int_0^{D(z)}d{\widetilde D}\,\eps_\0[\bx(\bq, {\widetilde D})]\;,
\label{ffa}\ee
where the integral is calculated along the trajectory of the fluid
element. Note that, since in the frozen-flow approximation
shell-crossing never occurs, the mapping $\bq \to \bx$ can be inverted
at any time. The solution (\ref{ffa}) might be replaced in
eq.(\ref{eq:biasris}) to obtain a non-local expression for the FFA bias
parameter. However, we can make a further step by noting that, for
Lagrangian points $\bq_\ast$ corresponding to local extrema of the
initial gravitational potential, $\nabla_{\bq}\varphi_\0(\bq_\ast)={\bf
0}$, FFA predicts $\bx_\ast=\bx(\bq_\ast,z)=\bq_\ast$, and
\be 
1+\de(\bx_\ast, z)=\exp\,\big[D(z)\,\eps_\0(\bx_\ast)\big]\;,
\ee
One can speculate that such points represent the preferential sites for
the formation of massive haloes, which could be associated to clusters
of galaxies, and use this approximate expression to obtain
\be
\rmb^E_{F\!F\!A}(\bx_\ast, z|M, z_f)\approx
1+\f{1+\de(\bx_\ast,z)}{\de(\bx_\ast,z)} \,\ln\big[1+\de(\bx_\ast,z)\big] 
\,\f{\rmb_\0^L(\bx_\ast|M,z_f)}{D(z)} \;.
\ee
Expanding this expression in powers of $\de$, to first-order we recover
the MW expression, eq.(\ref{eq:bhE}), while to second-order, we obtain
\be
b_{2F\!F\!A}^E (z|M,z_f) = \f{1}{D(z)^2\sigm^2} 
\left[ \f{\de_c^2}{D(z_f)^2\sigm^2}-3 \right] + 
\f{ D(z_f)}{D(z)} \left[\f{\de_c}{D(z_f)^2\sigm^2} - 
\f{1}{\de_c}\right] \;, 
\ee
which differs from the linear theory prediction of eq.(\ref{eq:bhE2}).
Analogous results could be obtained using the {\em frozen-potential}
approximation (Brainerd, Scherrer \& Villumsen, 1993; Bagla \&
Padmanabhan 1994), with the main difference that the $\de$ evolution
would be slowed down compared to FFA. Quite interesting is that the
{\em lognormal model} by Coles and Jones (1991) assumes that the
quantity $1+\de(\bx, z)$ can be always approximated by the exponential
of the linear density field at the same Eulerian position, so that the
expressions above for the Eulerian bias factor in FFA would apply to
all Eulerian points $\bx$. Of course, the validity of these approximate
expressions for the bias should be checked against the results of
N-body simulations.\\

Another way to get a local mapping between the evolved halo density
field and the underlying matter perturbations is to approximate the
non-linear evolution of the mass by the spherical top-hat model. This
method has been followed by MW and Mo \etal (1996).

\subsection{Perturbative evaluation of the Eulerian halo density contrast}

In the previous section we demonstrated that the Eulerian bias is a
non-linear and non-local function of the density fluctuation field. The
`non-locality', in particular, comes from the fact that the halo number
density fluctuation in $\bx$ is determined by the initial halo number
fluctuation at the Lagrangian position $\bq$, which, in turn, is
related to the linear mass fluctuation in the same point, through a
hierarchy of Lagrangian bias parameters. Here we want to derive an
approximate expression for $\de_h(\bx(\bq, D),D)$, by applying the
second-order Eulerian perturbation theory. Whenever it will be
necessary to go from the Lagrangian position $\bq$ to the Eulerian one,
the Zel'dovich approximation will show sufficient.

Within the linear regime, the Eulerian solution of the continuity
equation is simply, $\de^{(1)}(\bx, D) = D\,\eps_\0(\bx)$.  The mildly
non-linear regime may be approximately described by the second-order
solution (Bouchet \etal 1992; Bernardeau 1994; Catelan \etal 1995)
\be 
\de^{(2)}(\bx, D) =
\f{1}{2}\left[1-\f{E}{D^2}\right]\de^{(1)}(\bx, D)^2 
-D\,\bfu^{(1)}(\bx)\cdot\nabla\de^{(1)}(\bx, D) +
\f{1}{2}D^2\,\left[1+\f{E}{D^2}\right]\,
\p_{\al}u^{(1)}_{\beta}(\bx)\,\p_{\al}u^{(1)}_{\beta}(\bx)\;,
\label{7}
\ee
in such a way that $\de=\de^{(1)}+\de^{(2)}$ and higher-order
corrections are neglected. Here
$\bfu^{(1)}(\bx)=-\nabla\varphi_\0(\bx)$ is the (scaled) linear
peculiar velocity, and $\varphi_\0$ the (scaled) peculiar gravitational
potential, linearly extrapolated to the present time. The second-order
growth factor $E=E(D)$ is quite a complicated function of $D(\Om)$ (see
Appendix A, for its explicit expression), but, in the vicinity of $\Om
=1$ (actually in the range $0.05 \leq \Om \leq 3$), it can be
approximated by the expression $E \approx -\f{3}{7}\Om^{-2/63}D^2+
{\cal O}[(\Om - 1)^2]$ (see Bouchet \etal 1992). Therefore, the
previous second-order solution is well approximated by the expression
which holds in the Einstein de Sitter universe, namely (Fry 1984)
\be 
\de^{(2)}(\bx, D) =
\f{5}{7}\,\de^{(1)}(\bx, D)^2 
-D\,\bfu^{(1)}(\bx)\cdot\nabla\de^{(1)}(\bx, D) +
\f{2}{7}\,D^2
\p_{\al}u^{(1)}_{\beta}(\bx)\,\p_{\al}u^{(1)}_{\beta}(\bx)\;.
\label{8}
\ee

We want now to compute the corresponding second-order perturbative
correction, $\de_h^{(2)}(\bx, D)$, to the linear halo density
fluctuation field, $\de_h^{(1)}(\bfx, D)$. From equation
(\ref{eq:intcontin}) we obtain
\be
\de_h \approx \de^{(1)}_x + \de^{(2)}_x + b_1^L\,\de^{(1)}_q
+b_1^L \,\de^{(1)}_x\,\de^{(1)}_q+\f{1}{2}\, b_2^L\,\de^{(1)2}_q\;,
\ee
where, to maintain compact the notation we wrote
e.g. $\de^{(j)}_x\equiv \de^{(j)}(\bx, D)$. The Lagrangian bias factors
$b_1^L=b_1^L(z|M,z_f)$ and $b_2^L=b_2^L(z|M,z_f)$ are those given in
eqs.(\ref{eq:b1}) and (\ref{eq:b2}).  Notice that the perturbative
expansion of $\de_x$ holds at sufficiently early times and/or large
scales, while the validity of the expansion of $\de_h(\bq)$ in powers
of $\eps_\0(\bq)$ is based on assuming a suitably large smoothing
radius $R_\0$ on the background field $\eps(\bq)$.

The key point is that the first-order density field at the Lagrangian
position $\bq$ originates a non-local term, when written at the
corresponding Eulerian position $\bx$.  Using the Zel'dovich
approximation $\bx=\bq+D\,\bfu^{(1)}$, one obtains
$\de^{(1)}_q=\de^{(1)}_x-D\,\bfu^{(1)}\cdot\nabla\de^{(1)}_x$.
Finally, defining $\de_h=\de_h^{(1)}+\de_h^{(2)}$, one gets
$\de_h^{(1)}=(1+b^L_1)\,\de^{(1)}$ and
\be 
\de^{(2)}_h = 
\left[
\f{1}{2}\big(1-\f{E}{D^2}\big)+b_1^L + \f{1}{2}b_2^L\right]\,\de^{(1)2} 
-D\,\left(1+b_1^L\right)\,\bfu^{(1)}\cdot\nabla\de^{(1)} +
\f{1}{2}\,D^2\big(1+\f{E}{D^2}\big)\,
\p_{\al}u^{(1)}_{\beta}\,\p_{\al}u^{(1)}_{\beta}\;.
\label{77}
\ee
Thus, the non-locality has the effect of modifying the inertia term
$\bfu^{(1)}\cdot\nabla\de^{(1)}$, which gets multiplied by the factor
$(1+b_1^L)$.  The dynamical properties of the random field $\de_h$ may
be equivalently analyzed in terms of its Fourier transform $\fde_h(\bk,
t)$ where $\bk$ is the comoving wavevector. Thus, the second-order
solution (\ref{77}) may be written as
\be 
\fde^{(2)}_h(\bk, D) = D^2\int
\f{d\bk_1 d\bk_2}{(2\pi)^3}\, \ded(\bk_1+\bk_2-\bk)\,
\calH_S^{(2)}(\bk_1, \bk_2; b_1^L, b_2^L; \Om)\,
\fde_1(\bk_1)\,\fde_1(\bk_2)\;,
\label{12}
\ee
where the symmetrized kernel $\calH_S^{(2)}$ reads 
\be
\calH_S^{(2)}(\bk_1, \bk_2; b_1^L, b_2^L,\Om) \equiv 
\left[\f{1}{2}\Big(1-\f{E}{D^2}\Big)+b_1^L + \f{1}{2}b_2^L\right] +
\f{1+b_1^L}{2}
\left(\f{k_1}{k_2}+\f{k_2}{k_1}\right)\f{\bk_1\cdot\bk_2}{k_1\,k_2}
+\f{1}{2}\Big(1+\f{E}{D^2}\Big)
\left(\f{\bk_1\cdot\bk_2}{k_1\,k_2}\right)^2\;.
\label{14}
\ee 
The corresponding kernel for the Einstein-de Sitter case reads 
\be
\calH_S^{(2)}(\bk_1, \bk_2; b_1^L, b_2^L,\Om=1) \equiv 
\left[\f{5}{7} + b_1^L + \f{1}{2}b_2^L\right] +
\f{1+b_1^L}{2}
\left(\f{k_1}{k_2}+\f{k_2}{k_1}\right)\f{\bk_1\cdot\bk_2}{k_1\,k_2}
+\f{2}{7} \left(\f{\bk_1\cdot\bk_2}{k_1\,k_2}\right)^2\;.
\label{edsker}
\ee 

\subsection{Halo bispectrum and skewness}

A possible application of these results is the evaluation of the
bispectrum and corresponding skewness of the halo distribution.  A
related calculation has been performed by Fry (1996), who assumed the
bias to be local in Eulerian space at $z_f$.  It should be clear that
our model is quite different to the local Eulerian bias prescription
applied to the analysis of the skewness by Fry and Gazta\~naga (1993).
Moreover, the latter treatment, unlike ours, lacks of any prediction
for the value of the different bias parameters.  We recall that the
value of the gravitationally induced skewness of the mass is
\be
S=\f{\lan\de^3\ran}{\lan \de^2\ran^2}=4-2\f{E}{D^2}\;,
\label{skew1}
\ee
for unfiltered fields, and
\be
S({\cal R})=\f{\lan\de^3\ran}{\lan \de^2\ran^2}=4-2\f{E}{D^2}- 
\gamma({\cal R})\;,
\label{skew2}
\ee
for a spherical top-hat filter, where $\gamma\equiv -d\ln\s({\cal
R})^2/d\ln {\cal R}$ (Bernardeau 1994). The smoothing radius ${\cal R}$
should not be confused with $R$, defining the halo mass: one is
obviously interested in computing the skewness on a smoothing scale
much larger than the typical size of the single objects.  In the
Einstein-de Sitter universe, and for a scale-free power-spectrum, with
spectral index $n$, the latter reduces to $S({\cal R})=34/7-(n+3)$, for
$-3 \leq n<1$.

The derivation of the halo skewness $\lan\de_h^3\ran\approx
3\lan\de_h^{(1)2}\de_h^{(2)}\ran$ is simple. Assuming that the Eulerian
halo density field is smoothed by a top-hat filter, the halo skewness
parameter $S_h$ is, for a generic value of $\Om$,
\be
S_h({\cal R};z,\Om)=3\f{\lan\de_h^{(1)2}\,\de_h^{(2)}\ran}{\lan
\de_h^{(1)2}\ran^2}
=\f{4-2\f{E}{D^2}+6\,b_1^L(z|M,z_f)+3 \,b_2^L(z|M,z_f)-
\left[1+b_1^L(z|M,z_f)\right]\gamma({\cal R})}
{ \left[ 1+b_1^L(z|M,z_f) \right]^2}\;.
\ee
The asymptotic value of $S_h({\cal R};z,\Om)$, for a fixed formation
redshift $z_f$, is $S - \gamma({\cal R})$ as $z\to -1$, in the open and
flat cases, while, for $\Om>1$, this value is attained at the time of
maximum expansion, corresponding to $z=-1/\Om_\0$. This limit gives the
value of the underlying mass skewness: in the absence of merging the
haloes would evolve towards an unbiased distribution in the far
future. It is of interest to write the halo skewness in the Einstein-de
Sitter universe and for a scale-free linear power-spectrum,
\be
S_h(n;z,\Om=1)=\f{\f{34}{7}+6\,b_1^L(z|M,z_f)+3\,b_2^L(z|M,z_f)-(n+3)
\left[ 1+b_1^L(z|M,z_f) \right]}
{ \left[ 1+b_1^L(z|M,z_f) \right]^2}\;.
\ee
As for the mass skewness, the dependence on the smoothing scale ${\cal
R}$ now simply translates into a dependence on the spectral index
$n$. Again, the standard value for the mass skewness $34/7-(n+3)$ is
always recovered at the end of the expansion phase. The skewness
parameter is shown in Figure 3 for different values of $\Om_\0$ and for
a scale-free model with $n=-2$. For objects observed at the present
time, $z=0$, we vary the collapse epoch $z_f$, which may simulate
different models of galaxy formation inside dark haloes.  By varying
together $z=z_f$, we instead show the skewness evolution in the
instantaneous merging model. We also consider the case of varying only
$z$: this gives the evolution of the skewness in a model in which the
objects did not suffer any merging after their formation at $z_f$.
Finally, we show the evolution of the skewness parameter of filtered
mass fluctuations; note that the Einstein-de Sitter case displays no
redshift dependence, simply because of self-similarity; for sensible
values of $\Om_\0\neq 1$ also the mass skewness of non-flat Friedmann
models experiences very little evolution.  The redshift dependence of
$S_h$ is therefore mostly due to that of the Lagrangian bias
factors. Quite interesting, in this respect, is the fact that the halo
skewness plotted in the two top panels of Figure 3 displays a turning
point in its redshift dependence: this typically occurs when $M \approx
M_\ast(z_f)$.

\begin{figure}
\setlength{\unitlength}{1cm}
\centerline{\epsfxsize=8 cm \epsfbox{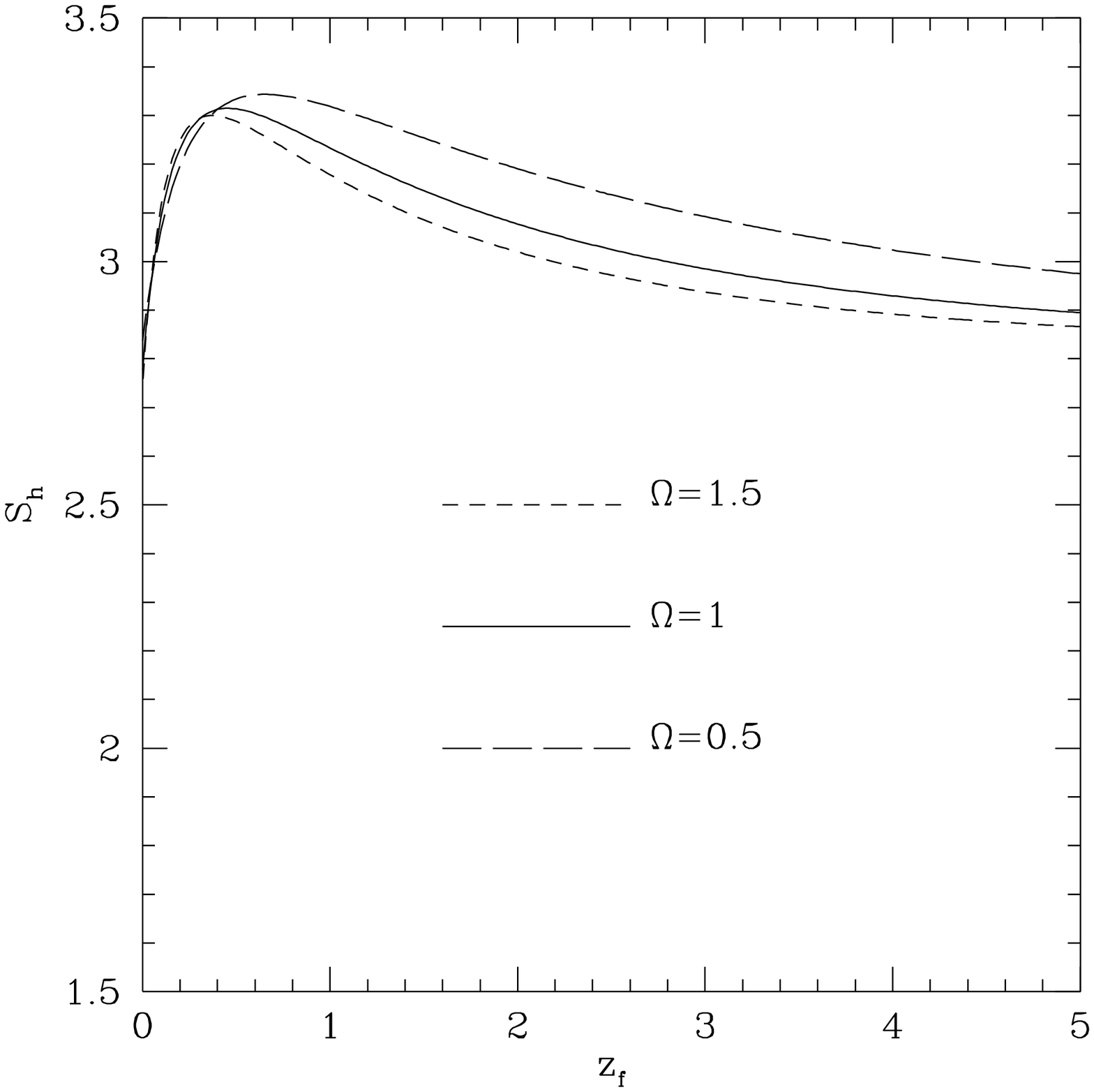}\epsfxsize=8 cm 
\epsfbox{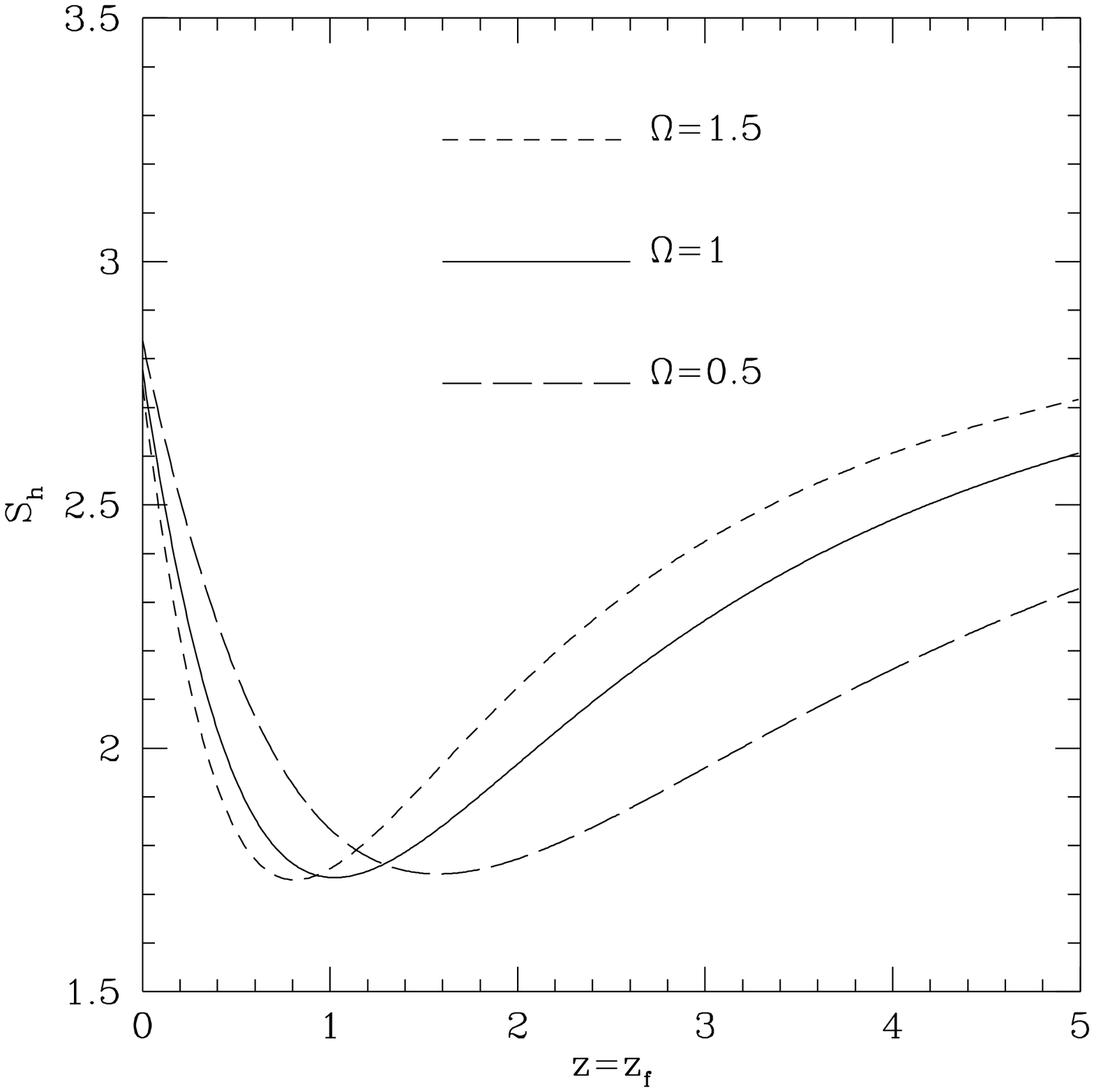}}
\centerline{\epsfxsize=8 cm \epsfbox{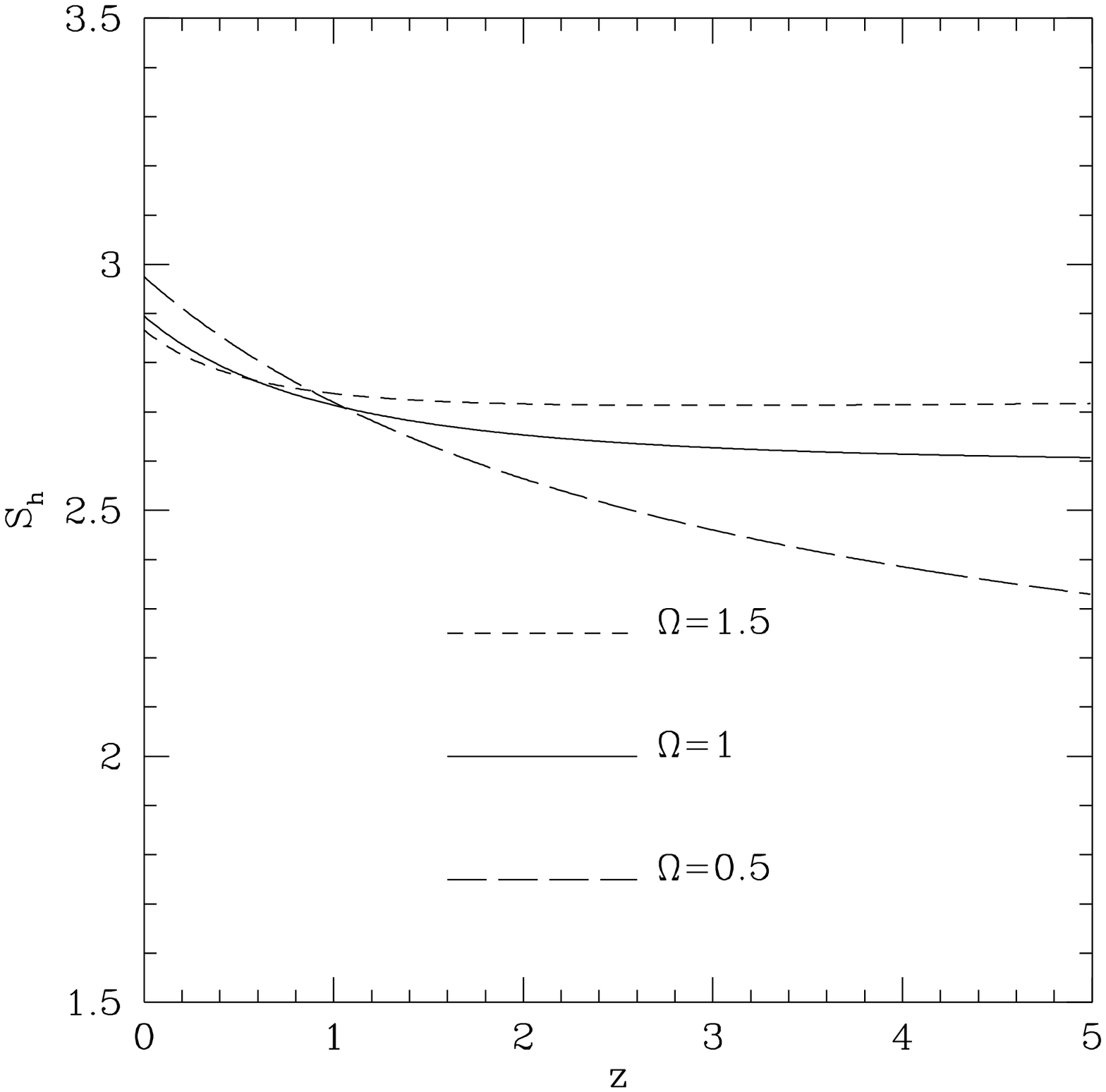}\epsfxsize=8 cm 
\epsfbox{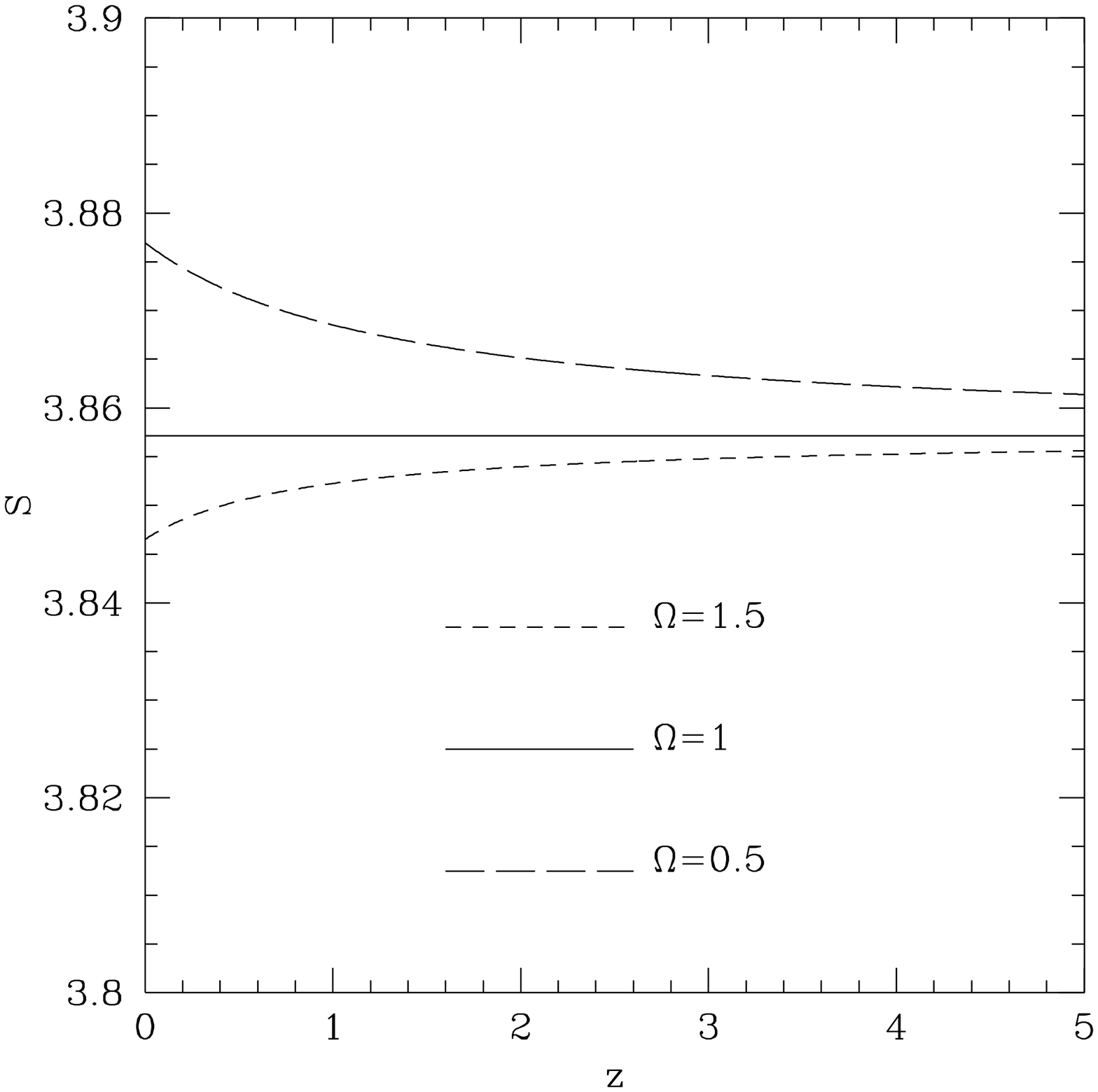}}
\caption{The filtered skewness parameter is plotted, for
$\Om_\0=0.5,~1,~1.5$, for a scale-free model with $n=-2$. The halo
masses are selected with the same linear mass variance $\sigm^2=10$,
corresponding to the same present-day bias parameters. We take
everywhere $\de_c=1.69$. The {\em top left} panel refers to objects
observed at $z=0$, with varying formation redshift $z_f$. The {\em top
right} panel shows the effect of varying simultaneously $z=z_f$. In the
{\em bottom left} panel we fix $z_f=5$ and look at different
observation redshifts $z\leq z_f$. The {\em bottom right} panel,
finally, shows the evolution of the skewness parameter of filtered mass
fluctuations.}
\label{fig:3}
\end{figure}

Of particular interest is also the expression for the halo bispectrum
$B_h$ defined by the relation
\be
\lan\fde_h(\bk_2, D)\,\fde_h(\bk_2, D)\,\fde_h(\bk_3, D)\ran \equiv 
(2\pi)^3\ded(\bk_1+\bk_2+\bk_3)\,B_h(\bk_1,\bk_2,\bk_3; D) \;. 
\ee
The leading term shows the characteristic hierarchical pattern
\be
B_h(\bk_1,\bk_2,\bk_3; D)= D^4\left[1+b_1^L(z|M,z_f)\right]^2
\left[ 2 ~ \calH_S^{(2)}(\bk_1, \bk_2; b_1^L, b_2^L,
\Om)\,P(k_1)\,P(k_2)\; 
+ \; 
{\rm cyclic\;\,terms} \; \right] \;, 
\ee
where $P(k)$ is the primordial density power-spectrum defined by
$\lan\fde_1(k_1)\,\fde_1(k_2)\ran=(2\pi)^3\ded(\bk_1+\bk_2)\,P(k_1)$,
and the two cyclic terms are obtained by the substitutions $\bk_1\to
\bk_2$, $\bk_1\to \bk_3$ and $\bk_2\to \bk_3$. Typically, as for the
hierarchical mass bispectrum, the halo bispectrum is largely scale
dependent, while its dependence on the $k$-shape is rather weak.  One
way to eliminate the scale dependence and look at the residual shape
dependence is to analyze the `effective' bispectrum amplitude $Q$ (Fry
1984)
\be Q\equiv \f{B_h(\bk_1,\bk_2,\bk_3; D) }
{P_h(k_1, D)\,P_h(k_2, D)+ P_h(k_1, D)\,P_h(k_2, D) + P_h(k_2,
D)\,P_h(k_3, D)}\;.
\ee
The halo power-spectrum is biased with respect to the mass one, $P_h(k,
D)=D^2\left[1+b_1^L(z)\right]^2 P(k)$.  For a power-law spectrum, the
amplitude $Q$ generally depends on the spectral index $n$, owing to the
wavenumber modulation introduced by the kernel
$\calH^{(2)}_S(\bk_1,\bk_2)$ (cf. Figure 4). For equilateral triangle
configurations, $Q$ gets an $n$-independent value, namely
\be
Q_{eq}(\Om; z)=
\f{
\f{1}{4}\left[1-3\,\f{E}{D^2}\right]+2\,b_1^L(z|M,z_f) + b_2^L(z|M,z_f)
}{\left[1+b_1^L(z|M,z_f)\right]^2}\;,
\ee
reducing to
\be
Q_{eq}(\Om=1; z)= \f{ \f{4}{7}+2\,b_1^L(z|M,z_f) + b_2^L(z|M,z_f)
}{\left[1+b_1^L(z|M,z_f)\right]^2}\;,
\ee
in the Einstein-de Sitter universe. 

\subsection{Local Lagrangian bias} 

So far, our model has been treated as being fully predictive. Once the
cosmological scenario and the structure formation model have been
fixed, our algorithm contains no fitting parameters. This is because we
used a local version of the PS theory to generate the Lagrangian halo
density contrast. One could, however, take a more general point of view
and assume that the Lagrangian halo density contrast $\de_h(\bq)$ is
specified in terms of the linear background density field
$\eps_{bg}(\bq,z) = D(z)\eps_\0(\bq)$ by a set of unknown bias
parameters ${\hat b}^L_\ell(z)$, as follows,
\be 
\de_h(\bq) = \sum_{\ell=1}^\infty \f{{\hat b}^L_{\0\ell}}{\ell!} 
\,\eps_\0(\bq)^\ell = 
\sum_{\ell=1}^\infty \f{{\hat b}^L_\ell(z)}{\ell!} 
\,\eps_{bg}(\bq,z)^\ell \;. 
\label{eq:loclagrbias}
\ee
Defining now $b_1 \equiv b_1(z) = 1 + {\hat b}^L_1(z)$ and $b_2 \equiv
b_2(z) = 2{\hat b}^L_1(z) + {\hat b}^L_2(z)$, according to
eq.(\ref{eq:itera}), and replacing these expansions in our previous
treatment, we recover the general expression (\ref{12}) for the
second-order halo density contrast, with the more general kernel
\be 
\calH_S^{(2)}(\bk_1, \bk_2; b_1, b_2,\Om) = 
\f{1}{2}\left[\Big(1-\f{E}{D^2}\Big)+ b_2 \right] +
\f{b_1}{2}
\left(\f{k_1}{k_2}+\f{k_2}{k_1}\right)\f{\bk_1\cdot\bk_2}{k_1\,k_2}
+\f{1}{2}\Big(1+\f{E}{D^2}\Big)
\left(\f{\bk_1\cdot\bk_2}{k_1\,k_2}\right)^2\;.
\label{eq:lockernel}
\ee
Comparing this relation with the analogous one obtained with a local
Eulerian bias expansion (e.g. Fry, Melott \& Shandarin 1995; Matarrese,
Verde \& Heavens 1997), we see that the bispectrum for a set of objects
selected by a local Lagrangian bias differs from the results of the
local Eulerian bias by the extra inertia term
\be 
\f{b_1-1}{2} 
\left(\f{k_1}{k_2}+\f{k_2}{k_1}\right)\f{\bk_1\cdot\bk_2}{k_1\,k_2} \;,
\ee 
which implies a different shape dependence. 

The halo bispectrum amplitude $Q(\theta)$, at $z=z_f=0$, for
configurations with sides $k_1=1$, $k_2=1/2$, separated by an angle
$\theta$, is shown in Figure 4, for scale-free models with $n=-2$ and
$n=-1$, with $\Om=1$.  Two different cases are considered: our local
Lagrangian bias model, with linear Eulerian parameters $b_1=2$ and
$b_2=1$ and the local Eulerian bias model of Fry and Gazta\~naga, with
the same Eulerian bias parameters.

\begin{figure}
\centerline{\epsfxsize= 8 cm \epsfbox{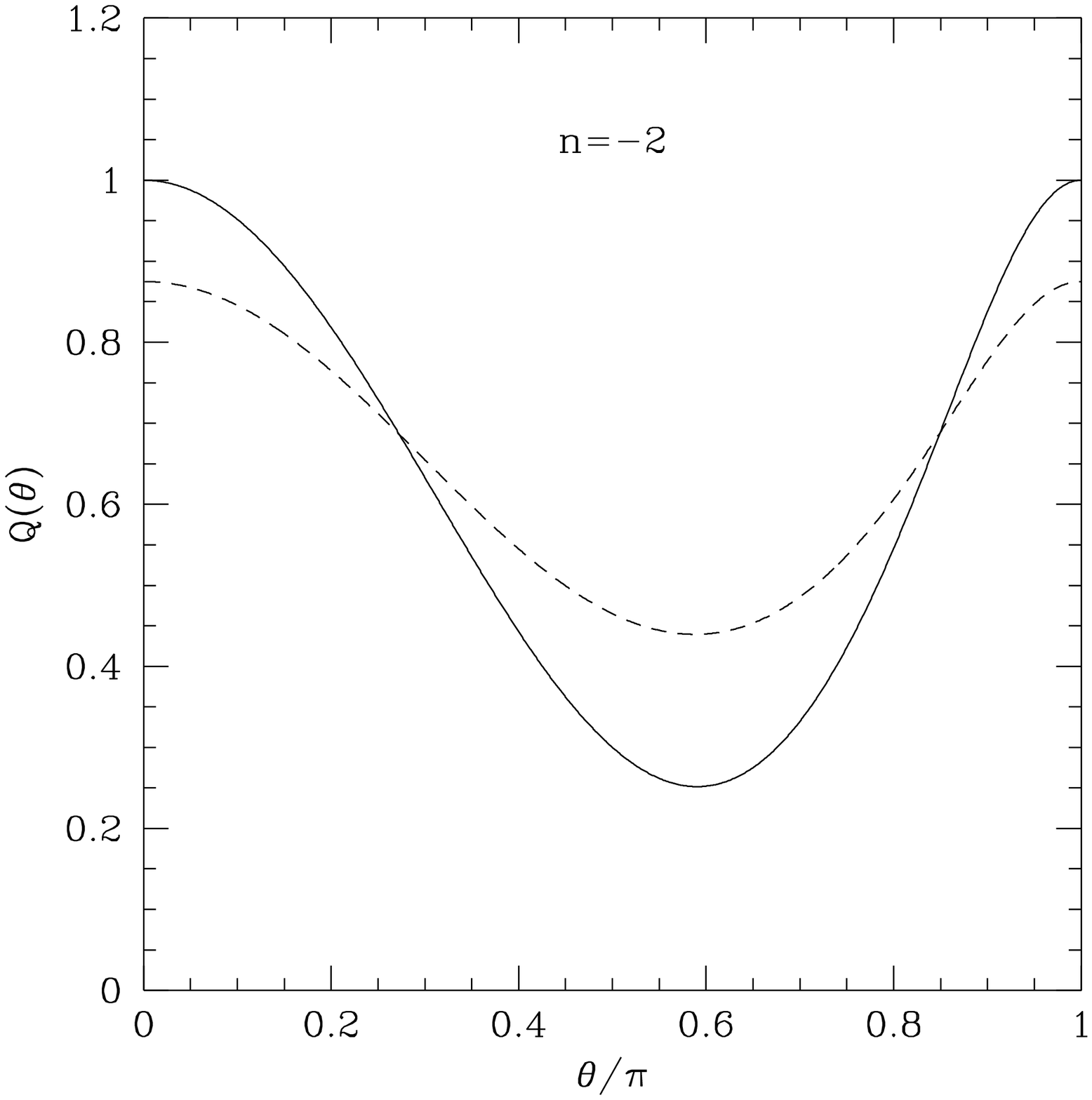}\epsfxsize= 8 cm 
\epsfbox{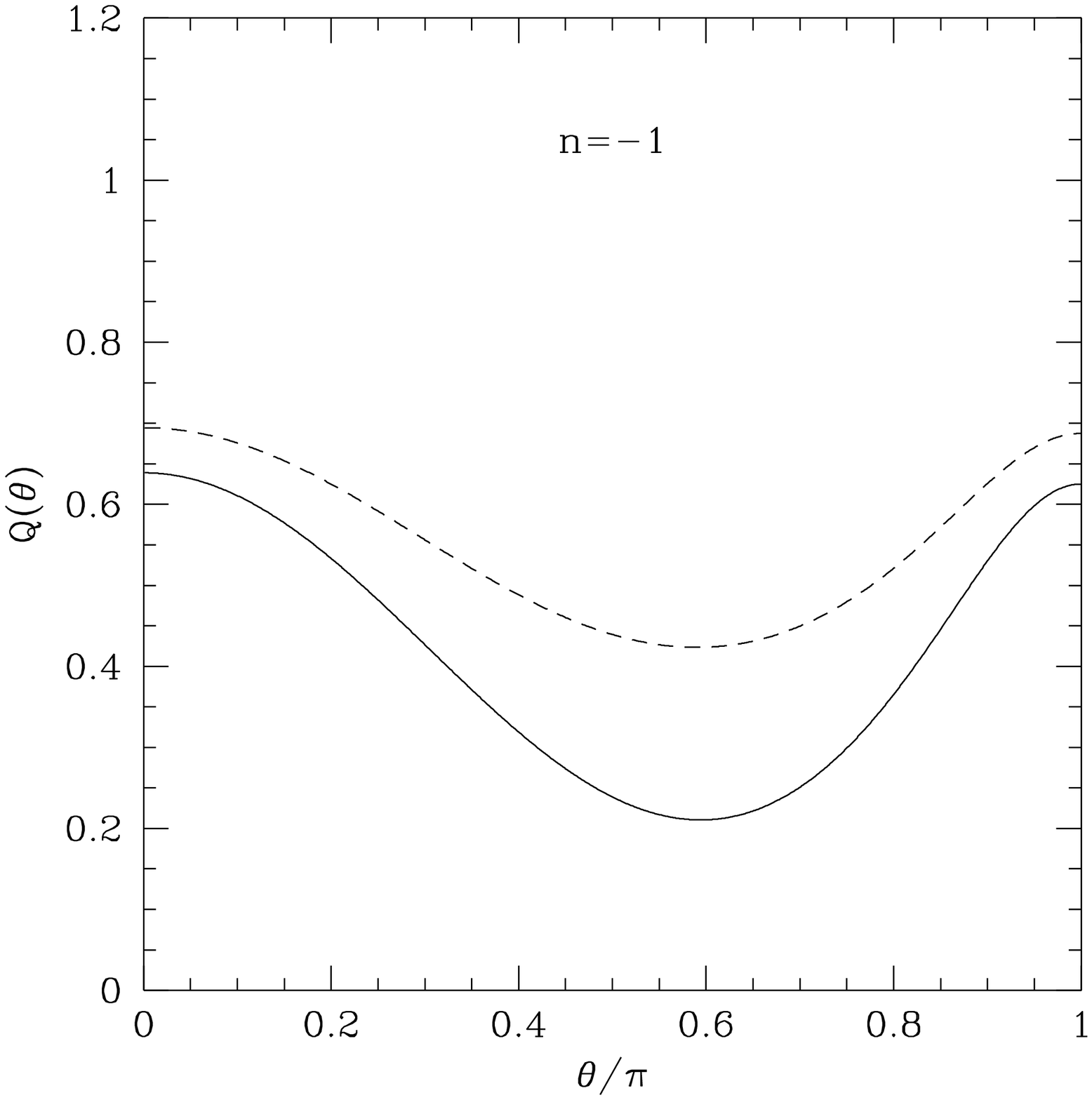}}
\caption{The halo bispectrum amplitude $Q(\theta)$ for configurations
with sides $k_1=1$, $k_2=1/2$, separated by an angle $\theta$ is
plotted vs.  $\theta$ for scale-free models with $n=-2$ and $n=-1$ at
$z=z_f=0$ and in a flat Universe. Two cases are shown for each panel:
the local Lagrangian bias model, with linear Eulerian parameters
$b_1=2$ and $b_2=1$ ({\em solid line}) and the local Eulerian bias
model, with the same bias parameters ({\em dashed line}).}
\label{fig:4}
\end{figure}

Similar reasoning would apply to the skewness, for which the local
Lagrangian vs. Eulerian bias hypothesis implies a change of the scale
dependence, through the extra term
\be 
- \,\f{b_1-1}{(b_1)^2}\, \gamma({\cal R}) \;. 
\ee 

\bigskip
\noindent With adequate modeling of galaxy formation inside dark matter
haloes (e.g. Kauffmann \etal 1997, and references therein) the results
of this section can be used to predict the clustering properties of
galaxies at different redshifts. In particular, the specific shape
dependence of the bispectrum (and related scale dependence of the
skewness), implied by our local Lagrangian bias prescription, would
reflect into a detectable signature in the statistical properties of
the galaxy distribution. Our model, therefore, provides a valid
alternative to local Eulerian bias schemes (e.g. Cen \& Ostriker 1992;
Coles 1993; Fry \& Gazta\~naga 1993; Catelan \etal 1994; Mann \etal
1997).

\section{Conclusions}

In this paper we studied the non-linear evolution of the clustering of
dark matter haloes, using a stochastic approach to single out the halo
formation sites directly in Lagrangian space. Our model is based on a
local version of the Press-Schechter theory, which becomes free of the
cloud-in-cloud problem after a suitable coarse-graining procedure is
applied.  The non-linear evolution of the halo distribution is then
followed exactly by relating it to the dynamics of the Lagrangian patch
of fluid which the nascent halo belongs to.

This formalism allowed us to obtain the bias random field relating the
local halo density contrast to the underlying mass distribution.  The
expression for the halo bias field, reported in eqs.(\ref{eq:biasris})
and (\ref{eq:biasjac}), represents the most relevant result of our
paper.  Because of the locality in Lagrangian space inherent in our
approach, such a bias field turns out to be non-local in Eulerian
coordinates, which has relevant implications for the clustering
properties of luminous objects like galaxies and galaxy clusters that
formed inside dark matter haloes.

Our method contains two Lagrangian smoothing scales. The scale $R$,
selecting the halo mass, and the background scale $R_\0 \gg R$ allowing
us to define the Lagrangian halo counting field as the local PS mass
function in a patch with comoving background density
$\varrho_b\,[1+\eps_{bg}(\bq,z)]$, $\eps_{bg}$ being the linear mass
fluctuation smoothed on the scale $R_\0$.  Given the role of the
latter, it would appear that our description of halo clustering makes
sense only on scales larger than $R_\0$. On the other hand, the
derivation of the Lagrangian correlation function in Section 2.4, which
does not make use of the background field, suggests that we can
actually extrapolate our Lagrangian results down to separation
comparable to the halo size.  This result is further confirmed by an
analysis in terms of space-correlated Langevin equations (Porciani
\etal 1997). The numerical results of MW and Mo \etal (1996) support
the idea that such an extrapolation would apply even in the
non-linearly evolved case.  In our treatment of the non-linear regime,
the background scale $R_\0$ appears with a complementary role.  It is
the minimum scale ensuring that the nascent haloes are indeed comoving
with the Lagrangian fluid patch which they belong to. This would
reasonably require that the Lagrangian fluid elements evolve with
negligible orbit crossing (e.g. Kofman \etal 1994).

Once again, let us stress that our approach makes no assumptions about
the merger rates of the considered objects. The clear distinction
between observation and formation redshift, $z$ and $z_f$, in our
approach implies that the instantaneous merging hypothesis, implicit in
the standard PS model, as well as any other realistic approximation can
be easily accommodated into our scheme as just the way to relate $z_f$
and $z$.

Our method for evolving the spatial distribution of the haloes is
indeed much more general than the specific application we have
considered so far. Given any Lagrangian population of objects specified
by some set of physical properties ${\cal M}$ (like mass and formation
threshold in our halo model), with conserved mean comoving number
density ${\bar n}_{obj}({\cal M})$ and local Lagrangian density
contrast $\de_{obj}(\bq|{\cal M})$, our results imply that, at any
redshift $z$, their comoving local density in Eulerian space is given
by
\be 
n_{obj} (\bx, z|{\cal M}) = {\bar n}_{obj}({\cal M}) 
\int d \bq \left[ 1 + \de_{obj}(\bq|{\cal M}) 
\right] \,\ded\!\left[\bx - \bx(\bq,z) \right] \;, 
\label{eq:convol}
\ee
where $\bx(\bq,z) = \bq + \bS(\bq,z)$, and $\bS(\bq,z)$ is the
displacement vector of the $\bq$-th Lagrangian element.  Smoothing the
initial gravitational potential on some scale $R_\0$ is again required,
so that the objects assigned to the $\bq$-th patch can be sensibly
assumed to be comoving with it.  This method could be used, for
instance, to follow the clustering of the Lagrangian density maxima in
the non-linearly evolved mass density field.  This suggests that,
starting from low-resolution numerical simulations, one can generate
mock catalogues of the given class of objects, with local density
correctly specified up to some resolution scale.  One can understand
the last relation as a local version of the Chapman-Kolmogorov equation
of stochastic processes (e.g. van Kampen 1992), stating that the local
Eulerian object distribution is the convolution of the Lagrangian
object density with the `conditional particle density', $\de_D [\bx -
\bx(\bq,z)]$, i.e. the probability that a particle is found in $\bx$ at
redshift $z$ given that it was in $\bq$ as $z\to \infty$.  The only
underlying hypothesis being, once again, that these objects move
exclusively by the action of gravity. It may be worth to notice that
the latter equation is actually more general than eq.
(\ref{eq:intcontin}), as it also holds in the presence of
multi-streaming.

Our non-linear stochastic approach can be already considered successful
in that, besides recovering the PS mass function, it provides a
self-consistent derivation of the Eulerian halo bias, which, to a first
approximation, reduces to the MW formula.  We, however, also predict
both quantitative and qualititive corrections to the MW results, that
clearly need to be checked against the outputs of numerical
simulations.  A definite prediction of our analysis is, for instance,
the form of the skewness and of the bispectrum of the spatial halo
distribution, which significantly deviates from that deduced with any
local Eulerian bias prescription.

\subsection*{Acknowledgements}
The authors greatly acknowledge Alan Heavens, Lev Kofman, Cedric Lacey
and Sergei Shandarin for many useful discussions.  PC is very grateful
to Eric Hivon. PC has been supported by the Danish National Research
Foundation at Copenhagen Theoretical Astrophysics Center (TAC), and by
the EEC HCMP CT930328 contract at Oxford Astrophysics Department, where
part of this investigation was performed. FL, SM and CP thank the
Italian MURST for partial financial support. CP is grateful to TAC for
kind hospitality.

\appendix
\section{Growth factors in Friedmann universe models}

The expressions for the first and second-order growth factors $D(z)$
and $E(z)$ have not been given in the main text. An easy derivation can
be given following Shandarin (1980) and using the relation
\be 
\Om^{-1} -1 = \left(\Om_\0^{-1} - 1 \right)(1+z)^{-1}\;.  
\ee 
We consider only cases with vanishing cosmological constant. The growth
factor ${\cal D} (z; \Om_\0)$ of linear density perturbations reads,
for the different geometries,
\begin{equation}
{\cal D}(z; \Om_\0)=\left\{
\begin{array}{ll}\f{5}{2} +\f{15}{2}\,\f{\Om_\0(1+z)}{1-\Om_\0}
\left[1-\f{1}{2}
\sqrt{\f{1+\Om_\0z}{1-\Om_\0}}
\,\ln\left(\f{\Om_\0(1+z)}{2-\Om_\0(1-z)-2\sqrt{(1-\Om_\0)(1+\Om_\0z) }}\right)
\right]
&\mbox{($\Om_\0<1$)}\\
(1+z)^{-1} & \mbox{($\Om_\0=1$)}\\
-\f{5}{2}+ \f{15}{2}\,
\f{\Om_\0(1+z)}{\Om_\0-1}
\left[
1+\sqrt{\f{1+\Om_\0z}{\Om_\0-1}}
\,\arctan\left(-\sqrt{\f {\Om_\0-1}  {1+\Om_\0z}}\,\right)
\right] &\mbox{($\Om_\0>1$)\,.}
\end{array}
\right.
\label{6}
\end{equation}
The expressions for the second-order growth factors ${\cal
E}(z;\Om_\0)$ are slightly more cumbersome:
\ba
{\cal E}(z; \Om_\0) = &-& \f{25}{8} -\f{225}{8}\,\f{\Om_\0(1+z)}{1-\Om_\0}\,
\left\{1-\f{1}{2}
\sqrt{\f{1+\Om_\0 z}{1-\Om_\0}}
\,\ln \Big(\f{\Om_\0(1+z)}{2-\Om_\0(1-z)-2\sqrt{(1-\Om_\0)(1+\Om_\0z) }}
\Big)
\right.
\nonumber \\
&+&
\left.
\f{1}{2}\left[-\sqrt{\f{1+\Om_\0z}{1-\Om_\0}} +
\f{1}{2}\f{\Om_\0(1+z)}{1-\Om_\0}\,
\,\ln\Big(\f{\Om_\0(1+z)}{2-\Om_\0(1-z)-2\sqrt{(1-\Om_\0)(1+\Om_\0z) }}\Big)
\right]^2 \right\} \;\;\;\mbox{($\Om_\0<1$)}\;,
\ea
\be
{\cal E}(z; \Om_\0) = -\f{3}{7(1+z)^2}\;\;\;\;\;\;\;\;\;\;\;\;\;\;\;\;\;
\mbox{($\Om_\0=1$)}\;,
\ee
\ba
{\cal E}(z; \Om_\0) = &-& \f{25}{8} +\f{225}{8}\,\f{\Om_\0(1+z)}{\Om_\0-1}\,
\left\{1+
\sqrt{\f{1+\Om_\0z}{\Om_\0-1}}
\,\arctan\left(-\sqrt{\f {\Om_\0-1}  {1+\Om_\0z}}\,\right) 
\right.
\nonumber \\
&+&
\left.
\f{1}{2}\left[
\sqrt{\f{1+\Om_\0z}{\Om_\0-1}} +
\f{\Om_\0(1+z)}{\Om_\0-1}
\,\arctan\left(-\sqrt{\f {\Om_\0-1}  {1+\Om_\0z}}\, \right)
\right]^2\right\}\;\;\;\;\;\;\;\;\mbox{($\Om_\0>1$)}\;.
\ea
Notice that we are implicitly adopting here the normalization suggested
by Shandarin (1980), so that, in the limit $z \to \infty$ one recovers
the Einstein de Sitter case, ${\cal D}(z; \Om_\0) \to
(1+z)^{-1}$. However, in the main text we normalized to unity the
linear growing factors extrapolated to the present time; so, for any
geometry, we define $D(z)\equiv {\cal D}(z; \Om_\0)/{\cal D}(z\!=\!0;
\Om_\0)$ and $E(z)\equiv {\cal E}(z; \Om_\0)/[{\cal D}(z\!=\!0;
\Om_\0)]^2$. \\

\end{document}